\documentclass[preprint,aps]{revtex4}
\usepackage{graphicx}
\usepackage{times}
\usepackage{latexsym}
\usepackage{amsmath}%
\usepackage{amsfonts}%
\usepackage{amssymb}

\begin{document}
\title{Electromagnetic response of a Gaussian beam to high-frequency
relic gravitational waves in quintessential inflationary models}
\renewcommand {\thefootnote} {\fnsymbol{footnote}}
\author{Fang-Yu Li$^{1,}$\footnote[1]{Email address: cqufangyuli@hotmail.com}\;\;\;\;\;Meng-Xi Tang$^2$\;\;\;\;\;Dong-Ping Shi$^1$}
\address {\footnotesize {1. Department of Physics, Chongqing University, Chongqing 400044,
People's Republic of China.\\
2. Department of Physics, Zhongshan University, Guangzhou 510275,
People's Republic of China.\\
\normalsize \rm{(revised manuscript, dated: 02-26-2003)}} }



\begin{abstract}
Maximal signal and peak of high-frequency relic gravitational
waves (GW's), recently expected by quintessential inflationary
models, may be firmly localized in the GHz region, the energy
density of the relic gravitons in critical units (i.e., $ h_0^2
\Omega _{GW}$) is of the order $10^{-6}$, roughly eight orders of
magnitude larger than in ordinary inflationary models. This is
just right best frequency band of the electromagnetic (EM)
response to the high-frequency GW's in smaller EM detecting
systems. We consider the EM response of a Gaussian beam passing
through a static magnetic field to a high-frequency relic GW. It
is found that under the synchroresonance condition, the
first-order perturbative EM power fluxes will contain ``left
circular wave" and ``right circular wave" around the symmetrical
axis of the Gaussian beam, but the perturbative effects produced
by the states of + polarization and $\times$ polarization of the
relic GW have different properties, and the perturbations on
behavior are obviously different from that of the background EM
fields in the local regions. For the high-frequency relic GW with
the typical parameters $ \nu _g = 10^{10}Hz$, $ h = 10^{ - 30} \ $
in the quintessential inflationary models, the corresponding
perturbative photon flux passing through the region $ \ 10^{ - 2}
m^{2} \ $ would be expected to be $ \ 10^{3}s^{-1} \ $. This is
largest perturbative photon flux we recently analyzed and
estimated using the typical laboratory parameters. In addition, we
also discuss geometrical phase shift generated by the
high-frequency relic
GW in the Gaussian beam and estimate possible physical effects.\\

PACS number(s): 04.30.Nk, 04.25.Nx, 04.30.Db, 04.80.Nn
\end{abstract}

\maketitle

\section{INTRODUCTION}

Relic GW's are very important sources of information on the very
early universe, physical behavior of the relic GW's expresses the
states and evolution of the early universe. Whether direct
detection or indirect tests to the relic GW's, both of them might
provide the new ways to observe our universe. On the other hand,
the expected properties of the relic GW's, such as amplitudes,
polarizations, frequency band, energy densities and spectra, et
al., are dependent on the concrete universe models. Thus the
expected features of the relic GW's and the concrete universe
models have a closed relation. In recent years, the quintessential
inflationary models have been much discussed \cite{1}-\cite{5},
and some astrophysical and cosmological observations seem to
indicate that \cite{1, 5, 6} the quintessential inflationary
models are explicit observationally acceptable. One important
expectation \cite{1, 2} of the models is that maximal signal and
peak of the relic GW's are firmly localized in the $GHz$ region,
the corresponding energy density of the relic gravitons is almost
eight orders of magnitude larger than in ordinary inflationary
models, and the dimensionless amplitude of the relic GW's in the
region can roughly reach up to $  10^{-30}  $ \cite{1}. This is
about five orders of magnitude more than that of the standing GW
discussed in Refs.\cite{7}-\cite{9}. Moreover, because the
resonant frequencies between the GW's and EM fields in some
smaller EM detecting systems (e. g., microwave cavities, strong EM
wave beams, and so on) [7-12], are just right distributed in the
GHz region, the results offered new hopes for the EM detection of
the GW's (Note that the frequency band detected by the VIRGO[13],
LIGO (Laser Interferometer Gravitational Wave Observatory)[14,
15], and LISA (Laser Interferometer Space Antenna)[16], are often
distributed in the region of $ \ 10^{-4} - 10^{4} \ $ Hz (this is
also most promising detecting frequency band for the usual
astronomical GW's), thus the EM response to the high-frequency
relic GW's might provide a new detecting window in the GHz band).

 In this paper, we shall study the EM response to the
high-frequency relic GW by a Gaussian beam propagating through a
static magnetic field. Here we consider it since there are
following reasons: (1)\;\;Unlike the usual EM response to the GW
by an ideal plane EM wave[17], the Gaussian beam is a realized EM
wave beam satisfying physical boundary conditions, and because of
the special property of the Gaussian function of the beam, the
resonant response of the Gaussian beam to the high-frequency GW's
has better space accumulation effect (see figure 5) than that of
the plane EM wave (see figure 7 in Ref.[8]). (2)\;\;In recent
years, strong and ultra-strong lasers and microwave beams are
being generated[18-21] under the laboratory conditions, and many
of the beams are often expressed as the Gaussian-type or the
quasi-Gaussian-type distribution, and they usually have good
monochromaticity in the GHz region. (3)\;\;The EM response in the
GHz band means that the dimensions of the EM system may reduced to
the typical laboratory size (e.g. magnitude of meter), thus the
requirements of other parameters can be further relaxed.
(4)\;\;Unlike the cavity electrodynamical response to the GW's (in
general, the detecting cavities are closed systems for the normal
EM modes stored inside the cavities), the Gaussian beam
propagating through a static EM field is an open system. In this
case the EM perturbations might have a more direct displaying
effect, although they have no energy accumulation effect in the
cavity electrodynamical response. Therefore, the EM response of
the Gaussian beams to the GW's and the EM detection of the
microwave cavities to the GW's have a very strong complementarity
each other, this is also one of motivations for this
investigation.

  The basic plan of this paper is the following. In Sec.II we
shall present usual form of the Gaussian beam in flat
space-time.In Sec.III we shall consider the EM response of the
Gaussian beam passing through a static magnetic field to the
high-frequency relic GW's in the quintessential inflationary
models. It includes the perturbation solutions of the
electrodynamical equations in curved space-time; the first-order
perturbative EM power fluxes (or in quantum language--perturbative
photon fluxes). Moreover, we shall give numerical estimations on
our results. In Sec.IV we will discuss possible geometrical phase
shift produced by the high-frequency relic GW's. Our conclusions
will be summarized in Sec.V.
\section{THE GAUSSIAN BEAM IN FLAT SPACE-TIME}
  It is well known that in flat space-time (i.e., when the GW's are
absent) usual form of the fundamental Gaussian beam is[22]\\
\begin{equation}
\psi=\frac{\psi_0}{\sqrt{1+(z/f)^{2}}}
\exp{(-\frac{r^{2}}{W^{2}})}\exp{\{i[(k_{e}z-
\omega_{e}t)-tan^{-1}{\frac{z}{f}}+\frac{k_{e}r}{2R}+\delta}]\},
\end{equation}\\
where $r^{2}=x^{2}+y^{2}$, $k_e=2\pi/\lambda_{e}$, $f=\pi
{W_0}^2/\lambda_e$, $W=W_0[1+(z/f)^2]^{\frac{1}{2}}$, $R=z+f^2/z$.
$\psi_0$
 is the maximal amplitude of the electric (or magnetic) field of
the Gaussian beam, i.e., the amplitude at the plane $z=0$, $W_0$
is the minimum spot size, namely, the spot radius at the plane
$z=0$, and $\delta$ is an arbitrary phase factor. $\psi$ satisfies
the scalar Helmholtz equation\\
\begin{equation}
\nabla^2\psi+{k_e}^2\psi=0.
\end{equation}\\
For the Gaussian beam in the vacuum, we have
${k_e}^2={\omega_e}^2\mu_0\epsilon_0$, where $\omega_e$ is the
angular frequency of the Gaussian beam.

Supposing that the electric field of the Gaussian beam is pointed
along the direction of the $x$-axis, that it is expressed as
Eq.(1), and a static magnetic field pointing
  along the $y$-axis is localized in the region $-l/2\leq z\leq
  l/2$, then we have

 {\[E^{(0)}={\tilde{E}_x}^{(0)}=\psi,\qquad
  E_y^{(0)}=E_z^{(0)}=0,\]

\begin{equation}
B^{(0)}  = \hat B^{(0)}  = \left\{ \begin{array}{l}
 \hat B_y^{(0)} {\;\;\;\;\;\;\;}( - {l \mathord{\left/
 {\vphantom {l {2 \le z \le {l \mathord{\left/
 {\vphantom {l 2}} \right.
 \kern-\nulldelimiterspace} 2}}}} \right.
 \kern-\nulldelimiterspace} {2 \le z \le {l \mathord{\left/
 {\vphantom {l 2}} \right.
 \kern-\nulldelimiterspace} 2}}}), \\
 0{\;\;\;\;\;\;\;\;\;\;\;\;}(z \le  - {l \mathord{\left/
 {\vphantom {l {2andz \ge {l \mathord{\left/
 {\vphantom {l 2}} \right.
 \kern-\nulldelimiterspace} 2}}}} \right.
 \kern-\nulldelimiterspace} {2{\;\;{\rm and}\;\;} z \ge {l \mathord{\left/
 {\vphantom {l 2}} \right.
 \kern-\nulldelimiterspace} 2}}}),
 \end{array} \right.
 \end{equation}\\
where the superscript 0 denotes the background EM fields, the
notations $\sim$ and $\wedge$ stand for the time-dependent and
static fields, respectively. Using (we use MKS units)\\
\begin{equation}
{\bf \tilde  B}^{(0)}  =  - \frac{i}{\omega _ e}\nabla  \times {
\bf \tilde E}^{(0)},
\end{equation}\\
and Eqs.(1) and (3), we obtain the time-dependent EM field
components in the cylindrical polar coordinates as follows
\begin{eqnarray}
\tilde {E}_r^{(0)}    & = & \psi \cos \phi,\qquad \tilde {E}_\phi
^{(0)}=-\psi \sin \phi,\qquad \tilde {E}_z^{(0)}=0,
\end{eqnarray}

\begin{eqnarray}
 \tilde {B}_r^{(0)}   & = & -\frac{i}{\omega _e} \frac {\partial \psi}{\partial z}\sin \phi \nonumber \\
                      & = & \left\{ {\frac{{\psi _0 \sin \phi }}{{\omega _e \left[ {1 + \left(
{{z \mathord{\left/
 {\vphantom {z f}} \right.
 \kern-\nulldelimiterspace} f}} \right)^2 } \right]^{1/2} }}} \right.
 \left[ {k_e  + \frac{{k_e r^2 \left( {f^2  - z^2 }
\right)}}{{2\left( {f^2  + z^2 } \right)^2}} - \frac{f}{{f^2  +
z^2
}}} \right] \nonumber \\
                      & + & \frac{i\psi _0 z \sin \phi}{\omega _e f^2
 [1+(z/f)^2]^{\frac{3}{2}}}\left. {\left[ {1 - \frac{{2r^2 }}{{W_0^2 \left[ {1 + \left( {{z
\mathord{\left/
 {\vphantom {z f}} \right.
 \kern-\nulldelimiterspace} f}} \right)^2 } \right]}}} \right]} \right\}\nonumber \\
                      &   & \cdot \exp{(-\frac{{r^2 }}{{W^2 }} )}\exp{\{i[ (k_e z-\omega _e
t)-tan^{-1}{\frac{z}{f}}+\frac{k_e
 r^2}{2R}+\delta ]\}},
 \end{eqnarray}

 \begin{eqnarray}
 \tilde {B}_\phi ^{(0)}
                      & = & -\frac{i}{\omega _e} \frac {\partial \psi}{\partial z}\cos \phi \nonumber \\
                      & = &
\left\{ {\frac{{\psi _0 \cos \phi }}{{\omega _e \left[ {1 + \left(
{{z \mathord{\left/
 {\vphantom {z f}} \right.
 \kern-\nulldelimiterspace} f}}
 \right)^2 } \right]^{1/2} }}} \right.
 \left[ {k_e  + \frac{{k_e r^2 \left( {f^2  - z^2 }
 \right)}}{{2\left( {f^2  + z^2 } \right)^2}} - \frac{f}{{f^2  + z^2 }}} \right]
 \nonumber \\
                      & + & \frac{i\psi _0 z \cos \phi}{\omega _e f^2
 [1+(z/f)^2]^{\frac{3}{2}}}\left. {\left[ {1 - \frac{{2r^2 }}{{W_0^2 \left[ {1 + \left( {{z
\mathord{\left/
 {\vphantom {z f}} \right.
 \kern-\nulldelimiterspace} f}} \right)^2 } \right]}}} \right]} \right\}\nonumber \\
                      &   & \cdot \exp{(-\frac{{r^2 }}{{W^2 }})}\exp{\{i[(k_e z-\omega _e t)-tan^{-1}{\frac{z}{f}}+\frac{k_e
 r^2}{2R}+\delta]\}},
 \end{eqnarray}

 \begin{eqnarray}
 \tilde {B}_z^{(0)}   & = & \frac{i}{\omega _e} \frac {\partial \psi}{\partial y}
 \nonumber \\
                      & = & -
\left\{ {\frac{{\psi _0 k_e r\sin \phi }}{{\omega _e \left[ {1 +
\left( {{z \mathord{\left/
 {\vphantom {z f}} \right.
 \kern-\nulldelimiterspace} f}} \right)^2 } \right]^{1/2} \left( {z + {{f^2 } \mathord{\left/
 {\vphantom {{f^2 } z}} \right.
 \kern-\nulldelimiterspace} z}} \right)}} + \frac{{i2\psi _0 r\sin \phi }}{{\omega _e W_0^2 \left[ {1 + \left( {{z \mathord{\left/
 {\vphantom {z f}} \right.
 \kern-\nulldelimiterspace} f}} \right)^2 } \right]^{3/2} }}} \right\}
\nonumber \\
                      &   & \cdot \exp{(-\frac{{r^2 }}{{W^2 }})}\exp{\{i[(k_e z-\omega _e t)-tan^{-1}{
\frac{z}{f} }+\frac{k_e
 r^2}{2R}+\delta]\}}.
\end{eqnarray}

  With the help of Eq.(1) and Eqs.(5)--(8), we can calculate the
power flux density of the Gaussian in flat space-time. For the
high-frequency EM power fluxes, only nonvanishing average values
of these with respect to time have an observable effect. From
Eqs.(1) and (5)-(8), one finds\\
\begin{eqnarray}
<\mathop {S^z }\limits^{\left( 0 \right)}>
 & = & \frac{1}{\mu_0}<\tilde{E}_x^{(0)} \tilde{B}_y^{(0)}>\nonumber
 \\ \nonumber
 & = & \frac {\psi _0 ^2}{2\mu_0\omega _e
 [1+(z/f)^2]}[k_e + \frac{k_e r^2
 (f^2-z^2)}{2(f^2+z^2)^2}-\frac{f}{f^2+z^2}]\exp{(-\frac{2r^2}{W
 ^2})},\\
 \\ \nonumber
<\mathop {S^r}\limits^{\left(0\right)}>
 & = & \frac{1}{\mu_0}<\tilde{E}_\phi^{(0)} \tilde{B}_z^{(0)}>\nonumber
 \\ \nonumber
 & = & \frac {\psi _0 ^2 k_e r\;sin^2\phi}{2\mu_0\omega _e
 [1+(z/f)^2](z+f^2/z)}\exp{(-\frac{2r^2}{W
 ^2})},\\
 \\ \nonumber
<\mathop {S^\phi}\limits^{\left(0\right)}>
 & = & -\frac{1}{\mu_0}<\tilde{E}_r^{(0)} \tilde{B}_z^{(0)}>\nonumber \\
 & = & \frac {\psi _0 ^2 k_e r\;\sin{(2\phi)}}{4\mu_0\omega _e
 [1+(z/f)^2](z+f^2/z)}\exp{(-\frac{2r^2}{W
 ^2})},
\end{eqnarray}\\
where $<\mathop {S^z }\limits^{\left( 0 \right)}>$, $<\mathop
{S^r}\limits^{\left(0\right)}>$ and $<\mathop
{S^\phi}\limits^{\left(0\right)}>$ represent the average values of
the axial, radial and tangential EM power flux densities,
respectively; the angular brackets denote the average values with
respect to time. We can see from Eqs.(9)-(11) that $ <\mathop
{S^r}\limits^{\left(0\right)}>_{z=0}\; = \;<\mathop
{S^\phi}\limits^{\left(0\right)}>_{z=0}\;\equiv\; 0,\;<\mathop
{S^z}\limits^{\left(0\right)}>_{z=0}\; =\; <\mathop
{S^z}\limits^{\left(0\right)}>_{max}\;$ and $\left| {<\mathop
{S^z}\limits^{\left(0\right)}>} \right| \; \gg \; \left|{<\mathop
{S^r}\limits^{\left(0\right)}>}\right|$\; and \;$\left| {<\mathop
{S^\phi}\limits^{\left(0\right)}>} \right|$\;in the region near
the minimum spot. Thus, the propagating direction of the Gaussian
beam is exactly parallel to the $z$-axis only in the plane $z=0$.
In the region of $z \neq 0$, because of nonvanishing $<\mathop
{S^r}\limits^{\left(0\right)}>$ and $<\mathop
{S^\phi}\limits^{\left(0\right)}>$ and the positive definite
property of $<\mathop {S^r}\limits^{\left(0\right)}>$, Eq.(10),
the Gaussian beam will be asymptotically spread as
$\left|{z}\right|$ increases.

  We will show that when the GW is present, the Gaussian beam will
be perturbed by the GW. Especially, under the synchroresonant
condition (i.e., when the frequency $\omega_g /2\pi$ of the GW
equals that $\omega_e /2\pi$ of the Gaussian beam), nonvanishing
first-order perturbative EM power fluxes can be produced, and they
on behavior are obviously different from that of the background EM
fields in the local regions.

\section{ELECTROMAGNETIC RESPONSE OF THE GAUSSIAN BEAM TO A HIGH-FREQUENCY RELIC GRAVITATIONAL WAVE}
\subsection {The high-frequency relic GW's in the quintessential inflationary models}

  For the relic graviton spectrum in the quintessential
inflationary models, recent analyses[1, 2] seem to indicate that
unlike the ordinary inflationary models, the maximum signal and
peak are firmly localized in the GHz region, the corresponding
energy density of the relic gravitons is almost eight orders of
magnitude larger than that in ordinary inflationary models, and
the dimensionless amplitude of the relic GW's in the GHz band can
roughly reach up to $10^{-30}$[1]. Thus smaller EM detecting
systems (not necessarily interferometers) may be suitable for the
detection purposes of the high-frequency relic GW's.

  In Fig.1 one illustrates the relic graviton logarithmic energy
spectra expected by the typical ordinary inflationary models(curve
I) and by the quintessential inflationary models (curve II),
respectively. Regions (1), (2), (3), (4)-1, (4)-2, (5) and (6)
represent the detecting frequency bands for the LISA[16], LIGO[14,
15], resonant-mass detectors[23-25], the superconducting microwave
cavities[10, 11, 38-40], the Gaussian beams tuned in on GHz
frequency band and Mini-ASTROD (Mini-Astrodynamical Space Test of
Relativity using Optical Devices) [26], respectively. Present
operating-mass detectors include ALLEGEO[23], EXPLORER[24], AURIGA
and NAUTILUS[25]. These detectors are often operating at KHz
frequency band. For example, the operating frequency of the
cryogenic resonant-mass detector EXPLORER is $923$ Hz[24]. The
sensitivity of them would be expected to be $\delta h \sim10^{ -
19} - 10^{ - 22}$ roughly for the GW's of KHz band.

\begin{figure}[htbp]
\centerline{\includegraphics[width=6.79in,height=3.50in]{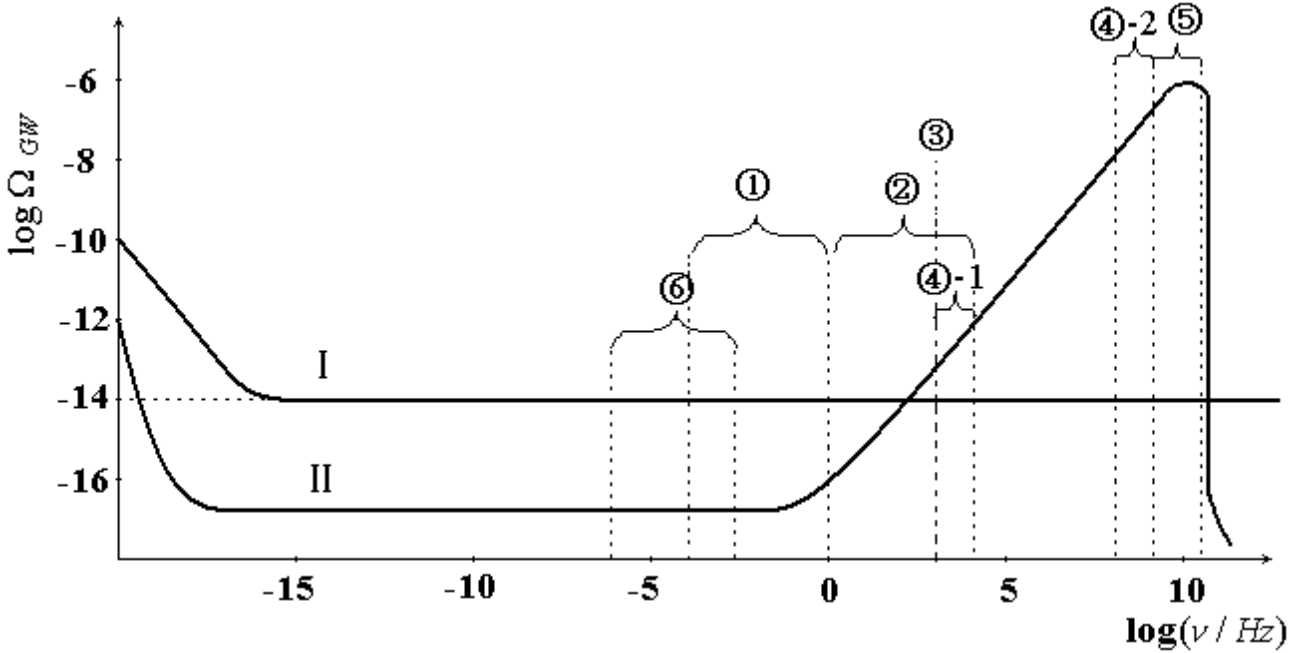}}
\caption{\footnotesize Curves I and II represent the relic
graviton logarithmic energy spectra (in critical units) expected
by the usual ordinary inflationary models and by the
quintessential inflationary models, respectively. The curves quote
from Ref.[1]. Here we illustrate roughly the distribution for some
detecting-frequency bands. The region (1) expresses the detecting
frequency band of LISA ($\sim{10^{-4}}-1 Hz $); the region (2) is
one of LIGO ($\sim {1} -10^4 Hz $); the region (3) is one of the
resonant-mass detectors ($\sim{10^3} Hz$); the region (4)-1 is one
of the difference-frequency resonant response of the microwave
cavities ($\sim 10^3-10^4Hz$), the region (4)-2 ($\sim
10^8-10^9Hz$) is one of the fundamental resonant response of the
microwave cavities; the region (5) is one of the Gaussian beams
($\sim{10^9} - 10^{11} Hz$) and region (6) is one of Mini-ASTROD
($\sim10^{-6}-10^{-3}Hz$). It can be seen that the region (4)-2
coincides partly with the maximal signal and peak of the relic
graviton energy spectra expected by the quintessential
inflationary models, the detecting frequency band (5) of the
Gaussian beam tuned in on GHz frequency region can be almost
completely localized in this peak value region. For the relevant
background in figure 1, we presented a brief introduction in
Appendixes (Appendix A: The cavity electromagnetic response to the
GW's. Appendix B: The dimensionless amplitude $h$ and the power
spectrum $S_h$ of the relic GW's. Appendix C: ASTROD. Appendix D:
Noise problems).}
\end{figure}

 Giovannini[1] analyzed the relic GW's in the quintessential
inflationary models within a framework of quantum theory. In this
framework the Fourier expansion of the field operators of the
relic GW's can be written as
\[\hat{\mu}_{\oplus} (\vec{x},\eta)=\frac{1}{(2\pi)^{3/2}}\int d^3 k_g [\hat{\mu}_{\oplus} (k_g,\eta)\;
e^{i\vec{k_g} . \vec{x}}+ \hat{\mu}_{\oplus}^{\ast}(k_g,\eta)\;
e^{-i\vec{k_g} . \vec{x}}],\nonumber\]
\begin{equation}
\hat{\mu}_{\otimes} (\vec{x},\eta)=\frac{1}{(2\pi)^{3/2}}\int d^3
k_g [\hat{\mu}_{\otimes} (k_g,\eta)\; e^{i\vec{k_g} . \vec{x}}+
\hat{\mu}_{\otimes}^{\ast}(k_g,\eta)\; e^{-i\vec{k_g} . \vec{x}}],
\end{equation}
where\\
\[\hat{\mu}_{\oplus} (k_g,\eta)=\psi_{\oplus}(k_g,\eta)\hat{a}_{\oplus}(k_g),\]
\begin{equation}
\hat{\mu}_{\otimes}(k_g,\eta)=\psi_{\otimes}(k_g,\eta)\hat{a}_{\otimes}(k_g),
\end{equation}\\
$\oplus$ and $\otimes$ denote + and $\times$ polarization of the
GW's, respectively, $\eta$ is the conformal time,
$\hat{a}_{\oplus}^*$ , $\hat{a}_{\otimes}^*$ and
$\hat{a}_{\oplus}$, $\hat{a}_{\otimes}$ represent the creation and
annihilation operators of the above two physical polarization
states. From the viewpoint of observation, the classical picture
of Eq.(12) corresponds to the amplitudes of the relic GW's with
the states of the two polarization, namely,
\[h_{\oplus}(\vec{x},\eta)=\frac{1}{(2\pi)^{3/2}}\int d^3 k_g \;h_{\oplus}(k_g,\eta)\; e^{i\vec{k_g} . \vec{x}},\nonumber\]
\begin{equation}
h_{\otimes}(\vec{x},\eta)=\frac{1}{(2\pi)^{3/2}}\int d^3 k_g \;
h_{\otimes}(k_g,\eta)\; e^{i\vec{k_g} . \vec{x}},
\end{equation}\\
where the integration spreads over the whole frequency band of the
relic GW's. However, if we hope to realize the resonant response
of a monochromatic Gaussian beam to the relic GW's, it is
necessary to let the frequency of the Gaussian beam equals a
certain frequency in the peak value region of the relic GW's
(i.e., $\omega_e = \omega_g$). Fortunately, most of strong
microwave beams (including the Gaussian beam) generated by present
technology are monochromatic or quasi-monochromatic, thus
resonateable relic GW to the Gaussian beam will be only one
monochromatic component of satisfying condition
$\omega_g=\omega_e$ in the relic GW frequency band. In this case
the corresponding treatment can be greatly simplified without
excluding the essential physical features. Of course, for the
resonant response to the relic GW's on the earth, we should use
interval of laboratory time (i.e., $c\;dt=a(\eta)\;d\eta$) and
laboratory frequency[27]. Consequently, a monochromatic circular
polarized plan relic GW propagating along the $z$-axis can be
written as\\
\[h_{\oplus}=h_{xx}=-h_{yy}=A_{\oplus}\;\exp({ik_{\alpha}x^{\alpha}})=A_{\oplus}\exp{[i(k_g z-\omega_g t)]},\nonumber\]
\begin{equation}
h_{\otimes}=h_{xy}=h_{yx}=iA_{\otimes}\;\exp({ik_{\alpha}x^{\alpha}})=iA_{\otimes}\exp{[i(k_g
z-\omega_g t)]}.
\end{equation}\\
This is just usual form of the GW in the TT gauge. Eq.(15) can be
viewed as the classical approximation of the Eqs.(12) and (13)
under the monochromatic wave condition.

  Since the relic GW's in the peak value region of the quintessential inflationary models have
very high frequency (the GHz frequency band), future measurements
through the microwave cavities or the strong Gaussian beams may be
useful. Especially, for the high-frequency relic GW's of $10^9 $
Hz $\leq \nu_g \leq 10^{11}$ Hz (these are just best peak
frequency band expected by the quintessential inflationary
models), using the EM response of the Gaussian beam would be more
suitable (see \vspace{1.5 mm} Fig.1).

\subsection {The electromagnetic system in the high-frequency relic gravitational wave field}

  From Eq.(15), the nonvanishing components of metric tensor in
Cartesian coordinates are given by\\
\[g_{00}=-1,\qquad g_{11}=g_{xx}=1+h_{xx}=1+h_{\oplus},\qquad g_{22}=g_{yy}=1+h_{yy}=1-h_{\oplus}, \nonumber\]
\[g_{12}=g_{21}=g_{xy}=g_{yx}=h_{xy}=h_{\otimes},\qquad g_{33}=g_{zz}=1,\nonumber\]
\[g^{00}=-1,\qquad g^{11}=g^{xx}=1-h^{xx}=1-h_{\oplus},\qquad g^{22}=g^{yy}=1-h^{yy}=1+h_{\oplus}, \nonumber\]
\begin{equation}
g^{12}=g^{21}=g^{xy}=g^{yx}=-h^{xy}=-h_{\otimes},\qquad
g^{33}=g^{zz}=1.
\end{equation}\\
  With the help of Eqs.(3)--(8) and considering perturbation
produced be the weak GW field expressed as Eqs.(15) and (16), the
components of the EM field tensor in the Cartesian coordinates can
be written as\\
\[F_{01}=\tilde{F}_{01}^{(0)}+\tilde{F}_{01}^{(1)}=\frac{1}{c}(\tilde{E}_{x}^{(0)}+\tilde{E}_{x}^{(1)})=
\frac{1}{c}(\psi + \tilde{E}_{x}^{(1)}),\nonumber\]
\[F_{02}=\tilde{F}_{02}^{(1)}=\frac{1}{c} \tilde{E}_{y}^{(1)},\qquad F_{03}=\tilde{F}_{03}^{(1)}=
\frac{1}{c}\tilde{E}_{z}^{(1)}, \nonumber\]
\[F_{12}=\tilde{F}_{12}^{(0)}+\tilde{F}_{12}^{(1)}=-(\tilde{B}_{z}^{(0)}+\tilde{B}_{z}^{(1)})=
-(\frac{i}{\omega}\frac{\partial{\psi}}{\partial{y}}+\tilde{B}_{z}^{(1)}),\nonumber\]
\[F_{13}=\hat{F}_{13}^{(0)}+ \tilde{F}_{13}^{(0)}+\tilde{F}_{13}^{(1)}=
\hat{B}_{y}^{(0)}+\tilde{B}_{y}^{(0)}+\tilde{B}_{y}^{(1)}=\hat{B}_{y}^{(0)}-\frac{i}{\omega}\frac{\partial{\psi}}{\partial{z}}+
\tilde{B}_{y}^{(1)}, \nonumber\]
\begin{equation}
F_{23}=\tilde{F}_{23}^{(1)}=\tilde{B}_{x}^{(1)},
\end{equation}\\
where $\hat{F}_{\mu\nu}^{(0)}$ and $\tilde{F}_{\mu\nu}^{(0)}$
represent the background static magnetic field $\hat{B}_y^{(0)}$
and the background EM wave field (the Gaussian beam),
respectively, $\tilde{F}_{\mu\nu}^{(1)}$ is the first-order
perturbation to the background EM fields in the presence of the
GW. For the nonvanishing $F_{\mu\nu}^{(0)}$ and
$F_{\mu\nu}^{(1)}$, we have
$\left|\tilde{F}_{\mu\nu}^{(1)}\right|\ll
\left|F_{\mu\nu}^{(0)}\right|$.

  The EM response to the GW can
be described by Maxwell equations in curved space-time, i.e.,\\
\begin{equation}
\frac{1}{\sqrt{-g}}\frac{\partial}{\partial{x^\nu}}(\sqrt{-g}\;
g^{\mu\alpha}g^{\nu\beta}F_{\alpha\beta})=\mu _0 J^\mu,
\end{equation}
\begin{equation}
F_{[\mu\nu,\alpha]}=0,
\end{equation}\\
where $J^\mu$ indicates the four-dimensional electric current
density. For the EM response in the vacuum, because it has neither
the real four-dimensional electric current nor the equivalent
electric current caused by the energy dissipation, such as Ohmic
losses in the cavity electrodynamical response or the dielectric
losses[10], so that $J^\mu=0\;$ in Eq.(18).

  Unlike the interaction of a plane EM wave with a plane GW (according
 to the Einstein-Maxwell equations under the condition
of the weak gravitational field, if both the plane GW and plane EM
wave have the same propagating direction, then the perturbation of
GW to the EM wave vanishes[17])., the electric and magnetic fields
of the Gaussian beam are nonsymmetric (see Eqs.(1) and (5)-(8)).
In this case, using Eqs.(5)-(8) and (18)-(19), it can be shown
that even for the plane GW propagating along the positive
$z$-direction, it can produce nonvanishing perturbative effect to
the Gaussian beam. In order to find the concrete form of the
perturbation produced by the direct interaction of the GW with the
Gaussian beam, it is necessary to solve Eqs.(18) and (19) by
substituting Eqs.(1), (5)-(8), (15), (16) and (17) into them,
which is often quite difficult. However, as shown in Refs.[8, 17,
28, 29], the orders of amplitudes of the first-order perturbative
electric and magnetic fields produced by the direct interaction of
the GW with the EM wave (e. g., plane wave or the Gaussian beam)
are approximately $ h\tilde B^{(0)} c\;$ and $h\tilde B^{(0)}$,
respectively, while that generated by the direct interaction of
the GW with the static magnetic field are approximately $ h\hat
B^{(0)} c \;$ and $h\hat B^{(0)}$ respectively. Thus corresponding
amplitude ratio is about $ h\tilde B^{(0)} /h\hat B^{(0)} $. In
our case, we have chosen $\tilde{B}^{(0)}\sim{10^{-3}}T$,
$\hat{B}^{(0)}\sim{10}T$, i.e., their ratio is only $10^{-4}$
roughly. Therefore, the former can be neglected. In other words,
the contribution of the Gaussian beam is mainly expressed as the
coherent synchroresonance (i.e., $\omega_e=\omega_g$) of it with
the first-order perturbation $\tilde{F}_{\mu\nu}^{(1)}$ generated
by the direct interaction of the GW with the static field
$\hat{B}_y^{(0)}$. In this case, solving the process of Eqs.(18)
and (19) can be greatly simplified, i.e., the static magnetic
field $\hat{B}^{(0)}=\hat{B}^{(0)}_y$ can be seen as the unique
background EM field in Eqs.(18) and (19). Under these
circumstances, in the first we can solve Eqs.(18) and (19) in the
region II ($-l /2\leq z \leq l/2, \;
\hat{B}^{(0)}=\hat{B}_y^{(0)}$) to find the first-order
perturbation solutions; and second, using the boundary conditions
one can obtain the first-order perturbation solutions in the
region I ($z \leq -l /2,\;\hat{B}^{(0)}=0$) and the region III
($z\geq l /2,\;\hat{B}^{(0)}=0$).

 Introducing Eqs.(15)-(17) into
Eqs.(18)-(19), neglecting high-order infinite small quantities and
perturbative effect produced by the direct interaction of the GW
with Gaussian beam, Eqs.(18) and (19) are reduced to\\
\[\frac{1}{c^2}\tilde{E}^{(1)}_{x,t}+\tilde{B}^{(1)}_{y,z}=\hat{B}_y^{(0)}h_{xx,z},\]
\begin{equation}
\tilde{E}^{(1)}_{x,z}+\tilde{B}^{(1)}_{y,t}=0,
\end{equation}\\
\[\frac{1}{c^2}\tilde{E}^{(1)}_{y,t}+\tilde{B}^{(1)}_{x,z}=\hat{B}_y^{(0)}h_{xy,z},\]
\begin{equation}
\tilde{E}^{(1)}_{y,z}-\tilde{B}^{(1)}_{x,t}=0,
\end{equation}
and
\begin{equation}
\tilde{E}^{(1)}_{z,t}=\tilde{E}^{(1)}_{z,z}=0,\qquad
\tilde{B}^{(1)}_{z,t}=\tilde{B}^{(1)}_{z,z}=0,
\end{equation}\\
where the commas denote partial derivatives. Using Eq.(15),
Eqs.(20)-(22) can also be expressed as following
inhomogeneous hyperbolic-type equations, respectively:\\
\[\square \tilde{E}_x^{(1)}            =  \tilde{E}_{x,zz}^{(1)}
-\frac{1}{c^2}\tilde{E}_{x,tt}^{(1)}  =
-A_{\oplus}\hat{B}_y^{(0)}k_g^2\,c\,\exp{[i(k_g z-\omega_g t)]},\]
\begin{equation}
\square \tilde{B}_y^{(1)}             =  \tilde{B}_{y,zz}^{(1)}
-\frac{1}{c^2}\tilde{B}_{y,tt}^{(1)}  =
-A_{\oplus}\hat{B}_y^{(0)}k_g^2\,\exp{[i(k_g z-\omega_g t)]},
\end{equation}

\[\square \tilde{E}_y^{(1)}             =  \tilde{E}_{y,zz}^{(1)}
-\frac{1}{c^2}\tilde{E}_{y,tt}^{(1)}  =
-iA_{\otimes}\hat{B}_y^{(0)}k_g^2\,c\,\exp{[i(k_g z-\omega_g
t)]},\]
\begin{equation}
\square \tilde{B}_x^{(1)}             =  \tilde{B}_{x,zz}^{(1)}
-\frac{1}{c^2}\tilde{B}_{x,tt}^{(1)}  =
iA_{\otimes}\hat{B}_y^{(0)}k_g^2\,\exp{[i(k_g z-\omega_g t)]},
\end{equation}

\begin{equation}
\square \tilde{E}_z^{(1)}   =  \square \tilde{B}_z^{(1)}=0,
\end{equation}\\
where $\square$ indicates the d'Alembertian. Obviously, every
solution of Eq.(20) must satisfy Eq.(23), and every solution of
Eq.(21) must satisfy Eq.(24), and it is easily seen from Eqs.(22)
and (25), that a physically reasonable solution of them would be
only

\begin{equation}
\tilde{E}_z^{(1)}=\tilde{B}_z^{(1)}=0.
\end{equation}\\

The general solutions of Eqs.(20)-(24) in the region II ($-l
/2\leq z
\leq l/2$) are given by\\
\[\tilde{E}_x^{(1)}=\frac{i}{2}A_{\oplus}\hat{B}_y^{(0)}k_g \,c \,(z+l/2)\exp{[i(k_g z-\omega_g t)]}+
b_1 \exp{[i(k_g z-\omega_g t)]}+c_1\exp{[i(k_g z+\omega_g t)]},\]
\begin{equation}
\tilde{B}_y^{(1)}=\frac{i}{2}A_{\oplus}\hat{B}_y^{(0)}k_g
(z+l/2)\exp{[i(k_g z-\omega_g t)]}+ b_2 \exp{[i(k_g z-\omega_g
t)]}+c_2\exp{[i(k_g z+\omega_g t)]},
\end{equation}

\[\tilde{E}_y^{(1)}=-\frac{1}{2}A_{\otimes}\hat{B}_y^{(0)}k_g
\,c\, (z+l/2)\exp{[i(k_g z-\omega_g t)]}+ ib_3 \exp{[i(k_g
z-\omega_g t)]}+ic_3\exp{[i(k_g z+\omega_g t)]},\]
\begin{equation}
\tilde{B}_x^{(1)}=\frac{1}{2}A_{\otimes}\hat{B}_y^{(0)}k_g
(z+l/2)\exp{[i(k_g z-\omega_g t)]}+ ib_4 \exp{[i(k_g z-\omega_g
t)]}+ic_4\exp{[i(k_g z+\omega_g t)]}.
\end{equation}\\
The solutions (26)-(28) show that perturbative EM waves produced
by the plane GW satisfing TT gauge, must be the transverse waves,
and the background static magnetic field is perpendicular to the
propagating direction of the GW. The constants in Eqs.(27) and
(28) satisfy\\
\[b_1-cb_2=-\frac{1}{2}A_{\oplus}\hat{B}_y^{(0)}c,\qquad c_1+cc_2=0,\]
\begin{equation}
b_3+cb_4=-\frac{1}{2}A_{\otimes}\hat{B}_y^{(0)}c,\qquad
c_3-cc_4=0,
\end{equation}\\
their concrete forms will be defined by the physical requirements
and boundary conditions.

The solutions (26)-(28) have the similar features as that found by
a number of authors [28, 29] previously, but as we shall show that
unlike the previous works, our EM system will be a possible scheme
to display the first-order EM perturbations produced by the
high-frequency relic GW's (GHz region) inside of the typical
laboratory size (not necessarily interferometers or EM cavities
with giant dimensions), a particularly interesting feature of the
first-order perturbation will be the perturbative effect in the
some special directions and in the some special regions.

It should be pointed out that in curved space-time, only local
measurements made by an observer travelling in his world-line have
definite observable meaning. These observable quantities are just
the projections of the physical quantities as the tensor on
tetrads of the observer's world-line. The tetrads consist of three
space-like mutually orthogonal vectors and a time-like vector
directed along the four-velocity of the observer, the latter is
perpendicular to the preceding ones. We indicate these with
$\tau^{\mu}_{(\alpha)}$, where the index in brackets numbers the
vectors and the other refers to the components of the tetrads in
the chosen coordinates. Consequently, the quantities
$F_{(\alpha\beta)} $ measured by the observer are the tetrad
components of the EM field tensor, that is
\begin{equation}
F_{(\alpha\beta)}=F_{\mu\nu}\tau^{\mu}_{(\alpha)}\tau^{\nu}_{(\beta)}.
\end{equation}
Obviously, for our EM system, the observer should be laid at rest
state to the static magnetic field, i.e., only the
zeroth-component of his four-velocity nonvanish. Thus, the tetrad
$\tau^{\mu}_{(0)}$ has the form
\begin{equation}
\tau^{\mu}_{(0)}=(\tau^0_{(0)},\;0,\;0,\;0).
\end{equation}

Using Eq.(16) and the orthonormality of the tetrads \;
$g_{\mu\nu}\tau^{\mu}_{(\alpha)}\tau^{\nu}_{(\beta)}=\eta_{\alpha\beta}$,\;
neglecting high-order infinite small quantities, it is always
possible to get
\begin{eqnarray}
\tau^\mu_{(0)} & = & (1,\;0,\; 0,\;0),\nonumber \\
\tau^\mu_{(1)} & = & (0,\;1-\frac{1}{2}h_{\oplus},\;0,\;0),\nonumber\\
\tau^\mu_{(2)} & = & (0,\;-h_{\otimes},\;1+\frac{1}{2}h_{\oplus},\;0),\nonumber\\
\tau^\mu_{(3)} & = & (0,\;0,\; 0,\;1).
\end{eqnarray}

Eq.(32) indicates that the zeroth and third components of the
tetrads coincide completely with the time- and $z$-axes in the
chosen coordinates, respectively. Furthermore, $\tau^\mu_{(1)}$
has only the projection on the $x$-axis, thus $\tau^\mu_{(1)}$
points actually at the $x$-axis. This means that the azimuth
$\phi$ in the tetrads and that in the chosen coordinates are the
same. While the deviation of $\tau^\mu_{(2)}$ from the $y$-axis is
only an order of $h_{\otimes}$.

With the help of Eqs.(3),(4),(27),(28) and (32), neglecting
high-order infinite small quantities, we obtain\\
\[E_{(x)}=cF_{(01)}=cF_{\mu\nu}\tau^\mu _{(0)}\tau^\nu _{(1)}=\tilde{E}_x^{(0)}+\tilde{E}_x^{(1)}-
\frac{1}{2}h_{xx}\tilde{E}_x^{(0)}=\psi+\tilde{E}_x^{(1)}-\frac{1}{2}h_{\oplus}\psi\;,\]
\[E_{(y)}=cF_{(02)}=cF_{\mu\nu}\tau^\mu_{(0)}\tau^\nu_{(2)}=\tilde{E}_y^{(1)}-h_{\otimes}\psi\;,\]
\[E_{(z)}=cF_{(03)}=cF_{\mu\nu}\tau^\mu_{(0)}\tau^\nu_{(3)}=0\;,\]
\[B_{(x)}=F_{(32)} = F_{\mu\nu}\tau^\mu _{(3)}\tau^\nu _{(2)}=\tilde{B}_x^{(1)}+h_{xy}(\hat{B}_y^{(0)}+
\tilde{B}_y^{(0)})=\tilde{B}_x^{(1)}+h_{\otimes}(\hat{B}_y^{(0)}-\frac{i}{\omega_e}\frac{\partial
\psi}{\partial z})\;,\]
\[B_{(y)}=F_{(13)}= F_{\mu\nu}\tau^\mu _{(1)}\tau^\nu _{(3)}=\hat{B}_y^{(0)}+\tilde{B}_y^{(0)}+
\tilde{B}_y^{(1)}=\hat{B}_y^{(0)}-\frac{i}{\omega_e}\frac{\partial
\psi}{\partial z}+\tilde{B}_y^{(1)}\;,\]
\begin{equation}
B_{(z)}=F_{(21)}=F_{\mu\nu}\tau^\mu_{(2)}\tau^\nu_{(1)}=\tilde{B}_z^{(0)}=\frac{i}{\omega_e}
\frac{\partial \psi}{\partial y}\;.
\end{equation}\\

As we have pointed out above, here
$|h\tilde{F}_{\mu\nu}^{(0)}|\sim
A\tilde{B}^{(0)}$\,,\,$|\tilde{F}_{\mu\nu}^{(1)}|\sim
A\hat{B}^{(0)}_y$\,,\, and in our case, $\hat{B}^{(0)}\sim
10T\,,\,\tilde{B}^{(0)}\sim 10^{-3}T\,,\,$ thus we have
$|h\tilde{F}_{\mu\nu}^{(0)}| / |\tilde{F}_{\mu\nu}^{(1)}|\;\approx
\tilde{B}^{(0)}/\hat{B}_y^{(0)} \approx 10^{-4}$\,,\, so that
\,$h\tilde{F}_{\mu\nu}^{(0)}$\, terms in Eq.(33) can be neglected
again. In this case Eq.(33) can be further reduced to (in the
region II: $-l/2\leq z\leq l/2$)
\begin{eqnarray}
E_{(x)}& = & \psi+\tilde{E}_x^{(1)}\nonumber\\
       & = &
       \psi+\frac{i}{2}A_{\oplus}\hat{B}_y^{(0)}k_g\,c\,(z+l/2)\exp{[i(k_g z-\omega_g
       t)]}+b_1\exp{[i(k_gz-\omega_gt)]}+c_1\exp{[i(k_gz+\omega_gt)]},\nonumber\\
B_{(y)}& = & \hat{B}_y^{(0)}-\frac{i}{\omega_e}\frac{\partial
\psi}{\partial z}+\tilde{B}_y^{(1)}\nonumber\\
       & = & \hat{B}_y^{(0)}-\frac{i}{\omega_e}\frac{\partial
\psi}{\partial
z}+\frac{i}{2}A_{\oplus}\hat{B}_y^{(0)}k_g(z+l/2)\exp{[i(k_g
z-\omega_g t)]}+b_2\exp{[i(k_gz-\omega_gt)]}\nonumber\\
       & + & c_2\exp{[i(k_gz+\omega_gt)]},
\end{eqnarray}
\begin{eqnarray}
E_{(y)}& = & \tilde{E}_y^{(1)}\nonumber\\
       & = &
       -\frac{1}{2}A_{\otimes}\hat{B}_y^{(0)}k_g\,c\,(z+l/2)\exp{[i(k_g z-\omega_g
       t)]}+ib_3\exp{[i(k_gz-\omega_gt)]}+ic_3\exp{[i(k_gz+\omega_gt)]},\nonumber\\
B_{(x)}& = & \tilde{B}_x^{(1)}+h_{\otimes}\hat{B}_y^{(0)}\nonumber\\
       & = & \frac{1}{2}A_{\otimes}\hat{B}_y^{(0)}k_g(z+l/2)\exp{[i(k_g
z-\omega_g
t)]}+ib_4\exp{[i(k_gz-\omega_gt)]}+ic_4\exp{[i(k_gz+\omega_gt)]}\nonumber\\
       &   & +iA_{\otimes}\hat{B}_y^{(0)}\exp{[i(k_gz-\omega_gt)]},
\end{eqnarray}
and \[E_{(z)}  =  0,\]
\begin{equation}
B_{(z)}  =  \tilde{B}_z^{(0)}=\frac{i}{\omega_e}\frac{\partial
\psi}{\partial y}\;\;_.
\end{equation}

\subsection {The particular solutions satisfying boundary
conditions}

  In fact, the perturbative parts in Eqs.(34)-(36) are the general solutions of Eqs.(20)-(25)
in the region II ($-l/2\leq z\leq l/2$). We shall define the
constants in Eqs.(34) and (35) to give the corresponding
particular solutions satisfying the boundary conditions.

  Clearly, the perturbative EM fields in the regions I, II and III
must satisfy the boundary conditions (the continuity conditions):
\begin{equation}
(\tilde{F}^{\;(1)}_{(\mu\nu)I})_{z=-l/2}\;\;=\;\;(\tilde{F}^{\;\,(1)}_{(\mu\nu)II})_{z=-l/2}\;,\qquad
(\tilde{F}^{\;\,(1)}_{(\mu\nu)II})_{z=l/2}\;\;=\;\;(\tilde{F}^{\;\;(1)}_{(\mu\nu)III})_{z=l/2}\;_.
\end{equation}

If one chooses the real part of the pure perturbative fields in
Eqs.(34) and (35), then we have
\begin{eqnarray}
\tilde{E}^{(1)}_{(x)}& = &
-\frac{1}{2}A_{\oplus}\hat{B}_y^{(0)}k_g\,c\,(z+l/2)\sin{(k_gz-\omega_g
t)}+
b_1\cos{(k_g z-\omega_g t)}+c_1\cos{(k_g z+\omega_g t)},\nonumber\\
\tilde{B}^{(1)}_{(y)}& = &
-\frac{1}{2}A_{\oplus}\hat{B}_y^{(0)}k_g(z+l/2)\sin{(k_g
z-\omega_g t)}+ b_2\cos{(k_g z-\omega_g t)}+c_2\cos{(k_g
z+\omega_g t)},
\end{eqnarray}

\begin{eqnarray}
 \tilde{E}^{(1)}_{(y)}& = &
-\frac{1}{2}A_{\otimes}\hat{B}_y^{(0)}k_g\,c\,(z+l/2)\cos{(k_gz-\omega_g
t)}-
b_3\sin{(k_g z-\omega_g t)}-c_3\sin{(k_g z+\omega_g t)},\nonumber\\
\tilde{B}^{(1)}_{(x)}& = &
\frac{1}{2}A_{\otimes}\hat{B}_y^{(0)}k_g(z+l/2)\cos{(k_g
z-\omega_g t)}- b_4\sin{(k_g z-\omega_g t)}-c_4\sin{(k_g
z+\omega_gt)}\nonumber\\
                     &   & -A_{\otimes}\hat{B}_y^{(0)}\sin{(k_gz-\omega_gt)}.
\end{eqnarray}

A physically reasonable requirement is that there is no the
perturbative EM wave propagating along the negative
 $z$-direction in the region III($z\geq l/2$). Here we shall
consider one more simple case, i.e., the perturbative EM wave in
the negative $z$-direction is also absent in the region
I($z\leq-l/2$). In order to satisfy the boundary conditions,
Eq.(37), and the above requirement, from Eqs.(29) and (37)-(39),
one finds
\[b_1  =  -\frac{1}{4}A_{\oplus}\hat{B}_y^{(0)}c,\qquad b_2
= \frac{1}{4}A_{\oplus}\hat{B}_y^{(0)},\qquad b_3 =
\frac{1}{4}A_{\otimes}\hat{B}_y^{(0)}c,\qquad b_4 =
-\frac{3}{4}A_{\otimes}\hat{B}_y^{(0)},\]
\[ c_1  =
\frac{1}{4}A_{\oplus}\hat{B}_y^{(0)}c,\qquad c_2  =
-\frac{1}{4}A_{\oplus}\hat{B}_y^{(0)},\qquad c_3=
\frac{1}{4}A_{\otimes}\hat{B}_y^{(0)}c,\qquad c_4 =
\frac{1}{4}A_{\otimes}\hat{B}_y^{(0)},\]
\begin{equation}
\end{equation}
and
\begin{equation}
l=n\lambda_g\;\;\;\;\;(n\;\;\rm {is\;\;integer}).
\end{equation}

In this case we have the perturbative EM fields in the above three
regions as follows:

 (a)\;\;Region I\;($z\leq -l/2$, $\hat{B}^{(0)}=0$)\\
\[\tilde{E}_{(x)}^{(1)}=c\tilde{F}_{(01)}^{(1)}=\tilde{E}_{(y)}^{(1)}=c\tilde{F}_{(02)}^{(1)}=0,\]
\begin{eqnarray}
\tilde{B}_{(x)}^{(1)}=\tilde{F}_{(32)}^{(1)}=\tilde{B}_{(y)}^{(1)}=\tilde{F}_{(13)}^{(1)}=0.
\\ \nonumber
\end{eqnarray}

 (b)\;\;Region II\;($-l/2\leq z\leq l/2$, $
\hat{B}^{(0)}=\hat{B}_y^{(0)}$)\\
\begin{eqnarray}
\tilde{E}_{(x)}^{(1)}&=&-\frac{1}{2}A_{\oplus}\hat{B}_y^{(0)}k_g\,c\,(z+l/2)\sin{(k_gz-\omega_gt)}
-\frac{1}{2}A_{\oplus}\hat{B}_y^{(0)}\,c\,\sin{(k_gz)}\sin{(\omega_gt)},\nonumber\\
\tilde{B}_{(y)}^{(1)}&=&-\frac{1}{2}A_{\oplus}\hat{B}_y^{(0)}k_g(z+l/2)\sin{(k_gz-\omega_gt)}
+\frac{1}{2}A_{\oplus}\hat{B}_y^{(0)}\sin{(k_gz)}\sin{(\omega_gt)},
\end{eqnarray}

\begin{eqnarray}
\tilde{E}_{(y)}^{(1)}&=&-\frac{1}{2}A_{\otimes}\hat{B}_y^{(0)}k_g\,c\,(z+l/2)\cos{(k_gz-\omega_gt)}
-\frac{1}{2}A_{\otimes}\hat{B}_y^{(0)}\,c\,\sin{(k_gz)}\cos{(\omega_gt)},\nonumber\\
\tilde{B}_{(x)}^{(1)}&=&\frac{1}{2}A_{\otimes}\hat{B}_y^{(0)}k_g(z+l/2)\cos{(k_gz-\omega_gt)}
-\frac{1}{2}A_{\otimes}\hat{B}_y^{(0)}\sin{(k_gz)}\cos{(\omega_gt)},
\end{eqnarray}

  (c)\;\;Region III\;($l/2\leq z\leq l_0, \hat{B}^{(0)}=0$)
\begin{eqnarray}
\tilde{E}_{(x)}^{(1)}&=&-\frac{1}{2}A_{\oplus}\hat{B}_y^{(0)}k_g\,c\,l
\,\sin{(k_gz-\omega_gt)},\nonumber\\
\tilde{B}_{(y)}^{(1)}&=&-\frac{1}{2}A_{\oplus}\hat{B}_y^{(0)}k_g\,l
\,\sin{(k_gz-\omega_gt)},
\end{eqnarray}
\begin{eqnarray}
\tilde{E}_{(y)}^{(1)}&=&-\frac{1}{2}A_{\otimes}\hat{B}_y^{(0)}k_g\,c\,l
\,\cos{(k_gz-\omega_gt)},\nonumber\\
\tilde{B}_{(x)}^{(1)}&=&\frac{1}{2}A_{\otimes}\hat{B}_y^{(0)}k_g\,l
\,\cos{(k_gz-\omega_gt)},
\end{eqnarray}
where $l_0$ is the size of the effective region in which the
second-order perturbative EM power fluxes, such as
$\frac{1}{\mu_0}(\tilde{E}_{(x)}^{(1)}\tilde{B}_{(y)}^{(1)})$,\;
$\frac{1}{\mu_0}(\tilde{E}_{(y)}^{(1)}\tilde{B}_{(x)}^{(1)})$,\;
keep a plane wave form. Notice that the power fluxes in the region
II contain the parts with space accumulation effect, i.e., they
depend upon the square of the interaction dimension. This is
because the GW's and EM waves have same velocity, so that the two
waves can generate an optimum coherent effect in the propagating
direction. It is easily to show that if we choose the imaginary
part of the pure perturbative fields in Eqs.(34) and (35), we can
obtain the similar results, but Eq.(41) will be replaced by\\
\begin{equation}
l=(2n+1)\frac{\lambda_g}{2}\;\;\;(n\;\;\rm {is\;\;integer}).
\end{equation}\\
Logi and Mickelson [29] used Feynmann perturbation techniques to
analyze the perturbative EM waves (photon fluxes) produced by a
weak GW (gravitons) passing through a static magnetic (or an
electrostatic) field, and found that perturbative EM waves (photon
fluxes) propagate only in the same and in the opposite propagating
directions of the GW(gravitons), the latter is weaker than the
former, or the latter is absent. Obviously, our results and
calculation by Logi et.al are self-consistent. However, due to the
weakness of the interaction of the GW's(gravitons) with the EM
fields(photons), we shall focus our attention on the first-order
perturbative power fluxes produced by the coherent
synchroresonance of the above perturbative EM fields with the
background Gaussian beam, not only the second-order perturbative
EM power fluxes themselves, such as
$\frac{1}{\mu_0}(\tilde{E}_{(x)}^{(1)}\tilde{B}_{(y)}^{(1)})$\;
and
$\frac{1}{\mu_0}(\tilde{E}_{(y)}^{(1)}\tilde{B}_{(x)}^{(1)})$.\;

\subsection {The first-order perturbative electromagnetic power
fluxes}

The generic expression of the energy-momentum tensor of the EM
fields in the GW fields is given by
\begin{eqnarray} T^{\mu \nu } =
\frac{1}{{\mu _0 }}\left( { - F_{\;\;{\rm    }\alpha }^\mu F^{\nu
\alpha }  + \frac{1}{4}g^{\mu \nu } F_{\alpha \beta } F^{\alpha
\beta } } \right).\nonumber\\
& &
\end{eqnarray}

Due to $ F_{\mu \nu }  = F_{\mu \nu }^{(0)}  + \tilde F_{\mu \nu
}^{(1)} $ and $ \left| {\tilde F_{\mu \nu }^{(1)} } \right| <  <
\left| {F_{\mu \nu }^{(0)} } \right| $ for the nonvanishing $
F_{\mu \nu }^{(0)} $ and $ \tilde F_{\mu \nu }^{(1)} $, $T^{\mu
\nu} $ can be disintegrated into\\
\begin{equation}
T^{\mu \nu }  = \mathop {T^{\mu \nu } }\limits^{(0)}  + \mathop
{T^{\mu \nu } }\limits^{(1)}  + \mathop {T^{\mu \nu }
}\limits^{(2)},
\end{equation}\\
where $ \mathop {T^{\mu \nu } }\limits^{(0)} $ is the
energy-momentum tensor of the background EM fields, and $\mathop
{T^{\mu \nu } }\limits^{(1)} $ and $\mathop {T^{\mu \nu }
}\limits^{(2)} $ represent first- and second-order perturbations
to $\mathop {T^{\mu \nu } }\limits^{(0)} $
 in the presence of the GW, respectively.

 Using Eq.(16), $\mathop {T^{^{\mu \nu } } }\limits^{(0)}$,
 $\mathop {T^{^{\mu \nu } } }\limits^{(1)}$ and $\mathop {T^{^{\mu \nu } } }\limits^{(2)}$
 can be written as
\begin{equation}
\mathop {T^{^{\mu \nu } } }\limits^{(0)}  = \frac{1}{{\mu _0
}}\left( { - F_{{\;\; \rm    }\alpha }^{\mu {\rm   }(0)} F^{\nu
\alpha (0)}  + \frac{1}{4}\eta ^{\mu \nu } F_{\alpha \beta }
^{(0)} F^{\alpha \beta (0)} } \right),
\end{equation}

\begin{eqnarray}
\mathop {T^{^{\mu \nu } } }\limits^{(1)}  = \frac{1}{{\mu _0
}}\left[ { - \left( {F_{{\;\; \rm    }\alpha }^{\mu {\rm   }(0)}
\tilde F^{\nu \alpha (1)}  + \tilde F_{{\;\; \rm    }\alpha }^{\mu
{\rm }(1)} F^{\nu \alpha (0)} } \right) + \frac{1}{4}\eta ^{\mu
\nu } \left( {\tilde F_{\alpha \beta } ^{(1)} F^{\alpha \beta (0)}
+ F_{\alpha \beta } ^{(0)} \tilde F^{\alpha \beta (1)} } \right) -
\frac{1}{4}h^{\mu \nu } F_{\alpha \beta } ^{(0)} F^{\alpha \beta
(0)} } \right],\nonumber\\
& &
\end{eqnarray}

\begin{eqnarray}
\mathop {T^{^{\mu \nu } } }\limits^{(2)}  = \frac{1}{{\mu _0
}}\left[ { - \tilde F_{\;\; {\rm    }\alpha }^{\mu {\rm   }(1)}
\tilde F^{\nu \alpha (1)} + \frac{1}{4}\eta ^{\mu \nu } \tilde
F_{\alpha \beta } ^{(1)} \tilde F^{\alpha \beta (1)}  -
\frac{1}{4}h^{\mu \nu } \left( {F_{\alpha \beta } ^{(0)} \tilde
F^{\alpha \beta (1)} + \tilde F_{\alpha \beta } ^{(1)} F^{\alpha
\beta (0)} } \right)} \right].
\end{eqnarray}
For the nonvanishing $ \mathop {T^{^{\mu \nu } } }\limits^{(0)} $,
$ \mathop {T^{^{\mu \nu } } }\limits^{(1)} $ and $ \mathop
{T^{^{\mu \nu } } }\limits^{(2)} $, we have
\begin{equation}
 | \mathop {T^{\mu \nu } }\limits^{(0)} | \gg | \mathop {T^{\mu
\nu } }\limits^{(1)} | \gg | \mathop {T^{\mu \nu } }\limits^{(2)}
| .
\end{equation}
 Therefore, for the effect of the
GW, we are interested in $ \mathop {T^{^{\mu \nu } }
}\limits^{(1)} $ but not in $ \mathop {T^{^{\mu \nu } }
}\limits^{(0)} $ and $ \mathop {T^{^{\mu \nu } } }\limits^{(2)} $.
Nevertheless, it can be shown from Eqs.(1), (3), (5)-(8),(15),
 (42)-(46) and (52), that the average value of $ \mathop {T^{^{00 }
} }\limits^{(2)} $ with respect to time is always positive. Thus
it expresses essentially the net increasing quantity of the energy
density of the EM fields. Especially, under the resonant
conditions, it should correspond to the resonant graviton-photon
conversion at a quantum level [29, 30, 31]. But because of $
|\mathop {T^{\mu \nu } }\limits^{(2)} | <  < |\mathop {T^{\mu \nu
} }\limits^{(1)} |
 $ for
the nonvanishing $ \mathop {T^{\mu \nu } }\limits^{(1)}
 $ and $ \mathop {T^{\mu \nu } }\limits^{(2)} $ , the second-order
perturbations are often far below the requirements of the
observable effect. In this case it has only theoretical interest.
However, for some astrophysical situations, it is possible to
cause observable effects, because where very large EM fields and
very strong GW's often occur simultaneously and these fields
extend over a very large area [32, 33].

By using
Eqs.(1), (3), (5)-(8), (15), (42)-(46) and (51). we obtain\\
\begin{eqnarray}
\mathop {S^r }\limits^{(1)}  = c\mathop {T^{01} }\limits^{(1)}  =
\frac{1}{{\mu _0 }}\left( {\tilde E_{(\phi)} ^{(1)} \tilde
B_{(z)}^{(0)} } \right) = - \frac{1}{{\mu _0 }}\left( {\tilde
E_{(x)}^{(1)} \tilde B_{(z)}^{(0)} } \right)\sin \phi  +
\frac{1}{{\mu _0 }}\left( {\tilde E_{(y)}^{(1)} \tilde
B_{(z)}^{(0)} } \right)\cos \phi,
\end{eqnarray}

\begin{eqnarray}
\mathop {S^\phi  }\limits^{(1)}  = c\mathop {T^{02} }\limits^{(1)}
=- \frac{1}{{\mu _0 }}\left( {\tilde E_{(r)}^{(1)} \tilde
B_{(z)}^{(0)} } \right) =  - \frac{1}{{\mu _0 }}\left( {\tilde
E_{(x)}^{(1)} \tilde B_{(z)}^{(0)} } \right)\cos \phi  -
\frac{1}{{\mu _0 }}\left( {\tilde E_{(y)}^{(1)} \tilde
B_{(z)}^{(0)} } \right)\sin \phi,
\end{eqnarray}

\begin{eqnarray}
\mathop {S^z }\limits^{(1)}  = c\mathop {T^{03} }\limits^{(1)}  =
\frac{1}{{\mu _0 }}\left( {\tilde E_{(x)}^{(0)} \tilde
B_{(y)}^{(1)} } \right) + \frac{1}{{\mu _0 }}\left( {\tilde
E_{(x)}^{(1)} \tilde B_{(y)}^{(0)} } \right).
\end{eqnarray}
where $ \mathop {S^r }\limits^{(1)} $ ,$ \mathop {S^\phi
}\limits^{(1)} $ and $ \mathop {S^z }\limits^{(1)} $ represent the
first-order radial, tangential and axial perturbative power flux
densities, respectively. As we have shown above, that for the
high-frequency perturbative power fluxes, only nonvanishing
average values of them with respect to time have observable
effect. It is easily seen from Eqs.(1), (5)-(8) and (42)-(46),
that average values of the perturbative power flux densities.
Eqs.(54)-(56),vanish in whole frequency rang of $\omega _ e \neq
\omega _ g$. In other words, only under the condition of $\omega _
e = \omega _ g$ (synchroresonance), $ \mathop {S^r }\limits^{(1)}
$ , $ \mathop {S^\phi }\limits^{(1)} $ and $ \mathop {S^z
}\limits^{(1)} $ have nonvanishing average values with respect to
time.

 In the following we study only the tangential average power
flux density $ < {\mathop {S^\phi  }\limits^{(1)} } > _{\omega _e
= \omega _g } $. Introducing Eqs.(1), (8), (42)-(46) into (55),
and setting $\delta=\pi/2$ in Eq.(1) (it is always possible), we
have
\begin{equation}
< {\mathop {S^\phi  }\limits^{(1)} } > _{\omega _e  = \omega _g }
= < {\mathop {S_ \oplus ^\phi }\limits^{(1)} } > _{\omega _e  =
\omega _g } + < {\mathop {S_ \otimes ^\phi }\limits^{(1)} } >
_{\omega _e  = \omega _g },
\end{equation}
where
$$
< {\mathop {S_ \oplus ^\phi  }\limits^{(1)} } > _{\omega _e  =
\omega _g }  =  - \frac{1}{{\mu _0 }}< {\tilde E_{(x)}^{(1)}
\tilde B_{(z)}^{(0)} } > \cos \phi  ,
$$
\begin{equation}
< {\mathop {S_ \otimes ^\phi  }\limits^{(1)} }
> _{\omega _e  = \omega _g }  =  - \frac{1}{{\mu _0
}}< {\tilde E_{(y)}^{(1)} \tilde B_{(z)}^{(0)} } > \sin \phi.
\end{equation}
$ < {\mathop {S_ \oplus ^\phi  }\limits^{(1)} }
> _{\omega _e  = \omega _g } $ and $ <
{\mathop {S_ \otimes ^\phi  }\limits^{(1)} } > _{\omega _e  =
\omega _g } $ represent the average values of the first-order
tangential perturbative power flux densities generated by the
states of $+$ polarization and $\times$ polarization of the GW,
Eq.(15), respectively. Using Eqs.(1), (8), (42)-(46), (57),
(58) and the boundary conditions [see, Eqs. (37) and (41)], one finds\\

 (a)\;\;Region I\;\;($z\leq-l/2$),

\begin{eqnarray}
< {\mathop {S_ \oplus ^\phi  }\limits^{(1)} }
> _{\omega _e  = \omega _g }  = < {\mathop
{S_ \otimes ^\phi  }\limits^{(1)} } > _{\omega _e  = \omega _g }
=0.
\\ \nonumber
\end{eqnarray}

(b)\;\;Region II\;\;($-l/2 \leq z \leq l/2$),

\begin{eqnarray}
 < {\mathop {S_ \oplus ^\phi  }\limits^{(1)} }
 > _{\omega _e  = \omega _g }  =
 \left\{ {\frac{{A_ \oplus  \hat B_y^{(0)} \psi _0 k_g r
 \left( {z + {l \mathord{\left/
 {\vphantom {l 2}} \right. \kern-\nulldelimiterspace} 2}}
 \right)}}{{8\mu _0 \left[ {1 + \left( {{z \mathord{\left/
 {\vphantom {z f}} \right. \kern-\nulldelimiterspace} f}}
\right)^2 } \right]^{1/2} \left( {z + {{f^2 } \mathord{\left/
 {\vphantom {{f^2 } z}} \right. \kern-\nulldelimiterspace} z}}
 \right)}}} \right.\cos \left( {\tan ^{ - 1} \frac{z}{f} -
 \frac{{k_g r^2 }}{{2R}}} \right)
\nonumber\\
 + \frac{{A_ \oplus  \hat B_y^{(0)} \psi _0 r\left( {z + {l \mathord{\left/
 {\vphantom {l 2}} \right.
 \kern-\nulldelimiterspace} 2}} \right)}}{{4\mu _0 W_0^2 \left[ {1 +
 \left( {{z \mathord{\left/
 {\vphantom {z f}} \right.
 \kern-\nulldelimiterspace} f}} \right)^2 } \right]^{3/2} }}\sin
 \left( {\tan ^{ - 1} \frac{z}{f} - \frac{{k_g r^2 }}{{2R}}} \right)
\nonumber \\
 - \frac{{A_ \oplus  \hat B_y^{(0)} \psi _0 r}}{{8\mu _0 \left[ {1 + \left( {{z \mathord{\left/
 {\vphantom {z f}} \right.
 \kern-\nulldelimiterspace} f}} \right)^2 } \right]^{1/2} \left( {z + {{f^2 } \mathord{\left/
 {\vphantom {{f^2 } z}} \right.
 \kern-\nulldelimiterspace} z}} \right)}}\sin \left( {k_g z} \right)\cos \left( {\tan ^{ - 1} \frac{z}{f} - \frac{{k_g r^2 }}{{2R}} - k_g z} \right)
\nonumber \\
 - \frac{{A_ \oplus  \hat B_y^{(0)} \psi _0 r}}{{4\mu _0 k_g W_0^2 \left[ {1 + \left( {{z \mathord{\left/
 {\vphantom {z f}} \right.
 \kern-\nulldelimiterspace} f}} \right)^2 } \right]^{3/2} }}\sin
 \left( {k_g z} \right)\sin \left. {\left( {\tan ^{ - 1} \frac{z}{f} -
 \frac{{k_g r^2 }}{{2R}} - k_g z} \right)} \right\} \exp(-\frac {r^2}{W^2}) \sin \left( {2\phi }
 \right),
\end{eqnarray}

\begin{eqnarray}
 < {\mathop {S_ \otimes ^\phi  }\limits^{(1)} } > _{\omega _e  = \omega _g }  = \left\{ {\frac{{A_ \otimes  \hat B_y^{(0)} \psi _0 k_g r\left( {z + {l \mathord{\left/
 {\vphantom {l 2}} \right.
 \kern-\nulldelimiterspace} 2}} \right)}}{{4\mu _0 \left[ {1 + \left( {{z \mathord{\left/
 {\vphantom {z f}} \right.
 \kern-\nulldelimiterspace} f}} \right)^2 } \right]^{1/2} \left( {z + {{f^2 } \mathord{\left/
 {\vphantom {{f^2 } z}} \right.
 \kern-\nulldelimiterspace} z}} \right)}}} \right.\sin \left( {\frac{{k_g r^2 }}{{2R}} -
  \tan ^{ - 1} \frac{z}{f}} \right)
\nonumber \\
  + \frac{{A_ \otimes  \hat B_y^{(0)} \psi _0 r\left( {z + {l \mathord{\left/
 {\vphantom {l 2}} \right.
 \kern-\nulldelimiterspace} 2}} \right)}}{{2\mu _0 W_0^2 \left[ {1 + \left( {{z \mathord{\left/
 {\vphantom {z f}} \right.
 \kern-\nulldelimiterspace} f}} \right)^2 } \right]^{3/2} }}\cos \left( {\frac{{k_g r^2 }}{{2R}} - \tan ^{ - 1} \frac{z}{f}} \right)
\nonumber \\
  + \frac{{A_ \otimes  \hat B_y^{(0)} \psi _0 r}}{{4\mu _0 \left[ {1 + \left( {{z \mathord{\left/
 {\vphantom {z f}} \right.
 \kern-\nulldelimiterspace} f}} \right)^2 } \right]^{1/2} \left( {z + {{f^2 } \mathord{\left/
 {\vphantom {{f^2 } z}} \right.
 \kern-\nulldelimiterspace} z}} \right)}}\sin \left( {k_g z} \right)
 \sin \left( {k_g z - \tan ^{ - 1} \frac{z}{f} + \frac{{k_g r^2 }}{{2R}} } \right)
 \nonumber \\
  + \frac{{A_ \otimes  \hat B_y^{(0)} \psi _0 r}}{{2\mu _0 k_g W_0^2 \left[ {1 + \left( {{z \mathord{\left/
 {\vphantom {z f}} \right.
 \kern-\nulldelimiterspace} f}} \right)^2 } \right]^{3/2} }}\sin
 \left( {k_g z} \right)\cos \left. {\left( {k_g z - \tan ^{ - 1}
 \frac{z}{f} + \frac{{k_g r^2 }}{{2R}} } \right)} \right\}\exp(-\frac {r^2}{W^2}) \sin ^2
 \phi.
 \\ \nonumber
\end{eqnarray}

(c)\;\;Region III\;\;($l/2 \leq z \leq l_0$),

\begin{eqnarray}
 < {\mathop {S_ \oplus ^\phi  }\limits^{(1)} } > _{\omega _e  = \omega _g }  = \left\{ {\frac{{A_ \oplus  \hat B_y^{(0)} \psi _0 k_g lr}}{{8\mu _0 \left[ {1 + \left( {{z \mathord{\left/
 {\vphantom {z f}} \right.
 \kern-\nulldelimiterspace} f}} \right)^2 } \right]^{1/2} \left( {z + {{f^2 } \mathord{\left/
 {\vphantom {{f^2 } z}} \right.
 \kern-\nulldelimiterspace} z}} \right)}}} \right.\cos \left( {\tan ^{ - 1} \frac{z}{f} - \frac{{k_g r^2 }}{{2R}}} \right)
\nonumber \\
  + \frac{{A_ \oplus  \hat B_y^{(0)} \psi _0 lr}}{{4\mu _0 W_0^2 \left[ {1 + \left( {{z \mathord{\left/
 {\vphantom {z f}} \right.
 \kern-\nulldelimiterspace} f}} \right)^2 } \right]^{3/2} }}\sin \left. {\left( {\tan ^{ - 1}
 \frac{z}{f} - \frac{{k_g r^2 }}{{2R}}} \right)} \right\}\exp(-\frac {r^2}{W^2}) \sin \left( {2\phi }
 \right),
\end{eqnarray}

\begin{eqnarray}
 < {\mathop {S_ \otimes ^\phi  }\limits^{(1)} } > _{\omega _e  = \omega _g }  = \left\{ {\frac{{A_ \otimes  \hat B_y^{(0)} \psi _0 k_g lr}}{{4\mu _0 \left[ {1 + \left( {{z \mathord{\left/
 {\vphantom {z f}} \right.
 \kern-\nulldelimiterspace} f}} \right)^2 } \right]^{1/2} \left( {z + {{f^2 } \mathord{\left/
 {\vphantom {{f^2 } z}} \right.
 \kern-\nulldelimiterspace} z}} \right)}}} \right.\sin \left( {\frac{{k_g r^2 }}{{2R}} - \tan ^{ - 1} \frac{z}{f}} \right)
\nonumber \\
  + \frac{{A_ \otimes  \hat B_y^{(0)} \psi _0 lr}}{{2\mu _0 W_0^2 \left[ {1 + \left( {{z \mathord{\left/
 {\vphantom {z f}} \right.
 \kern-\nulldelimiterspace} f}} \right)^2 } \right]^{3/2} }}\cos \left.
 {\left( {\frac{{k_g r^2 }}{{2R}} - \tan ^{ - 1} \frac{z}{f}} \right)} \right\}\exp(-\frac {r^2}{W^2}) \sin ^2
 \phi.
 \\ \nonumber
\end{eqnarray}

It is easily shown that the nonvanishing first-order perturbative
power flux densities are much greater than corresponding
second-order perturbative power flux densities. The quantum
picture of this process can be described as the interaction of the
photons with the gravitons in a background of virtual photons (or
virtual gravitons) as a catalyst [29, 34], which can greatly
increase the interacting cross section between the photons and
gravitons. In other words the interaction may effectively change
the physical behavior (e.g., propagating direction, distribution,
polarization and phase) of the photons in the local regions, even
if the net increasing quantities of the photon number (the EM
energy) of the entire EM system approach zero, such properties may
be very useful to display very weak signals of the GW's.

Eqs. (60)-(63) show that because there are nonvanishing $ {\mathop
{S_ \oplus ^\phi  }\limits^{(1)} } $ (which depends on the $+$
polarization state of the GW) and $ {\mathop {S_ \otimes ^\phi
}\limits^{(1)} } $ (which depends on the $\times$ polarization
state of the GW), the first-order tangential perturbative power
fluxes are expressed as the ``left circular wave" and ``right
circular wave" in the cylindrical polar coordinates, which around
the symmetrical axis of the Gaussian beam,  but $ < {\mathop {S_
\oplus ^\phi  }\limits^{(1)} } > _{\omega _e  = \omega _g } $ and
$ < {\mathop {S_ \otimes ^\phi }\limits^{(1)} } > _{\omega _e  =
\omega _g } $ have a different physical behavior. By comparing
Eqs.(60)-(63) with Eqs.(9)-(11), we can see that: (a)$ < {\mathop
{S_ \oplus ^\phi }\limits^{(1)} } > _{\omega _e  = \omega _g } $
and $ < {\mathop {S^\phi }\limits^{(0)} } > $
 have the same angular distribution factor $sin(2\phi)$ , thus $ < {\mathop {S_
\oplus ^\phi  }\limits^{(1)} } > _{\omega _e  = \omega _g } $ will
be swamped by the background power flux $ < {\mathop {S^\phi
}\limits^{(0)} }
> $. Namely, in this case $ < {\mathop {S_
\oplus ^\phi  }\limits^{(1)} } > _{\omega _e  = \omega _g } $ has
no observable effect. (b) The angular distribution factor of $ <
{\mathop {S_ \otimes ^\phi }\limits^{(1)} }
> _{\omega _e  = \omega _g } $ is a $sin^2 \phi$ , it
is different from that of $
 < {\mathop {S^\phi  }\limits^{(0)} }
> $. Therefore,$ < {\mathop {S_ \otimes
^\phi  }\limits^{(1)} } > _{\omega _e  = \omega _g } $,in
principle, has an observable effect. In particular, at the
surfaces $\phi = \pi /2, 3\pi /2$, $ < {\mathop {S^\phi
}\limits^{(0)} }
> \equiv 0 $, while $ | {< {\mathop
{S_ \otimes ^\phi  }\limits^{(1)} } > _{\omega _e  = \omega _g } }
| = | {< {\mathop {S_ \otimes ^\phi }\limits^{(1)} } > _{\omega _e
= \omega _g } } |_{\max } $ , this is satisfactory (although $ <
{\mathop {S^r }\limits^{(0)} } > $ and $ {< {\mathop {S_ \otimes
^\phi }\limits^{(1)} } > _{\omega _e  = \omega _g } } $ have the
same angular distribution factor $sin^2 \phi$, the propagating
direction of $ < {\mathop {S^r }\limits^{(0)} }
> $ is perpendicular to that of $<
\mathop {S_ \otimes ^\phi  }\limits^{(1)} >_{\omega _ e = \omega _
g}$ , thus $<\mathop {S ^r }\limits^{(0)}>$ (including $ <
{\mathop {S^z }\limits^{(0)} }
> $ ) has no any essentially contribution in the pure tangential
direction).

Figure 2 gives the distribution of $ < {\mathop {S_ \oplus ^\phi
}\limits^{(1)} } > _{\omega _e  = \omega _g } $ at the plane $
z=l/2= \frac{n}{2} \lambda _ g $ (n is a integer) in the
cylindrical polar coordinates, while figure 3 gives the
distribution of $ < {\mathop {S_ \otimes ^\phi  }\limits^{(1)} }
> _{\omega _e  = \omega _g } $ at the plane $z=0$ in
the cylindrical polar coordinates.

\begin{figure}[htbp]
\centerline{\includegraphics[width=5.05in,height=3.11in]{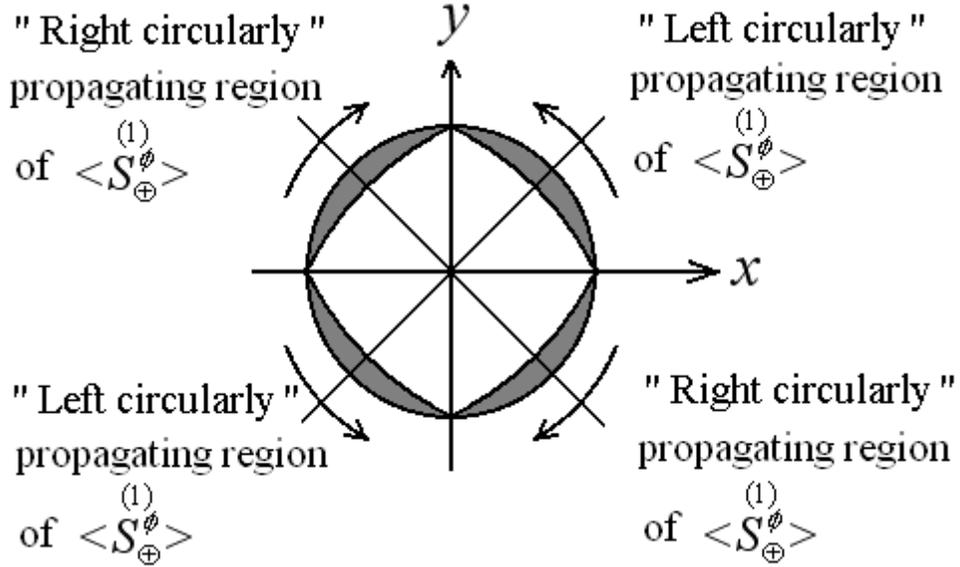}}
\caption{\footnotesize Distribution of $ < {\mathop {S_ \oplus
^\phi }\limits^{(1)} } > _{\omega _e  = \omega _g } $ at the plane
$ z=l/2= \frac{n}{2} \lambda _ g $ (n is integer) in the
cylindrical polar coordinates. It has maximum at $ \phi =\pi
/4,3\pi /4, 5 \pi /4,$ and $ 7 \pi /4 $, while it vanishes at
$\phi = 0, \pi /2,\pi$ and
 $3\pi/2$, here $l=0.1m$,$\lambda _ g = 0.01m$ and $r=0.05m$, and
the GW propagates along the $z$-axis. } \label{Vr_orbit1_part}
\end{figure}

\begin{figure}[htbp]
\centerline{\includegraphics[width=2.42in,height=2.20in]{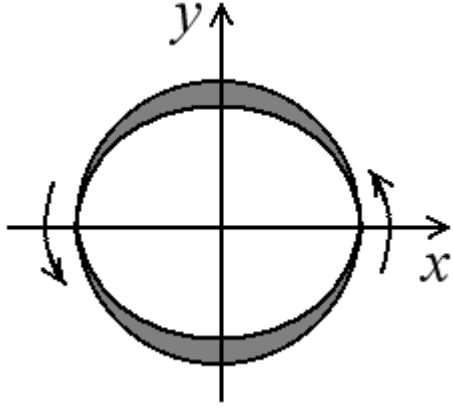}}
\caption{\footnotesize Distribution of $ < {\mathop {S_ \otimes
^\phi }\limits^{(1)} } > _{\omega _e  = \omega _g } $ at the plane
$ z=0 $  in the cylindrical polar coordinates. It has maximum at $
\phi =\pi /2$ and $3\pi /2 $,while it vanishes at $\phi = 0, \pi$,
Unlike figure 2, $ < {\mathop {S_ \otimes ^\phi }\limits^{(1)} } >
_{\omega _e  = \omega _g } $ at the $z=0$ is completely
``left-hand circular",  here $r=0.05m$, $\lambda _ g = 0.01m$, and
the GW propagates along the $z$-axis. } \label{Vr_orbit1_part}
\end{figure}

From Eqs.(10),(11),(60) and (61), we can also see that $ <
{\mathop {S^r }\limits^{(0)} } >  = < {\mathop {S^\phi
}\limits^{(0)} } >  = < {\mathop {S_ \oplus ^\phi }\limits^{(1)} }
> _{\omega _e  = \omega _g } \equiv 0 $ at the plane $z=0$, while
where
\begin{equation}
< {\mathop {S_ \otimes ^\phi  }\limits^{(1)} }
> _{\omega _e  = \omega _g }  = \frac{{A_ \otimes \hat
B_y^{(0)} \psi _0 lr}}{{4\mu _0 W_0^2 }} \exp \left( { -
\frac{{r^2 }}{{W_0^2 }}} \right)  \sin ^2 \phi.
\end{equation}

In Table I we list the distribution of $ < \mathop {S^\phi
}\limits^{(0)}  > $ , $ < \mathop {S_ \oplus ^\phi  }\limits^{(1)}
>_{\omega _e  = \omega _g } $  and $ < \mathop {S_ \otimes ^\phi  }\limits^{(1)}  > _{\omega _e  = \omega _g }
$ in some typical regions.

\begin{table}[!htbp] 
\scriptsize
 \caption {The distribution of $ < {\mathop
{S^\phi }\limits^{(0)} } > $,$ < {\mathop {S_ \oplus ^\phi
}\limits^{(1)} } > _{\omega _e  = \omega _g } $and $ < {\mathop
{S_ \otimes ^\phi }\limits^{(1)} }
> _{\omega _e  = \omega _g } $ in some typical
regions}

\begin{center}
\begin{tabular}{lccccc}\hline
 { } & { Angular distribution} & {The plane } & {The planes} &
{The planes} & {The planes} \\
{ } & {factor} & { $z=0$} & {$\phi = \pi /4, 5\pi /4$} & {$\phi =
3
\pi /4, 7 \pi /4$} & {$\phi = \pi /2, 3 \pi /2$}\\ \hline  \\
 $
< {\mathop {S^\phi  }\limits^{(0)} } > $
 & $sin(2\phi)$ & 0  & $ | {< {\mathop {S^\phi
}\limits^{(0)} } > } | = | {< {\mathop {S^\phi }\limits^{(0)} }
> } |_{\max } $, & $ | {< {\mathop {S^\phi
}\limits^{(0)} } > } | = | {< {\mathop {S^\phi }\limits^{(0)} }
> } |_{\max },
$  & 0  \\
 & & & ``Left circularly"  & ``Right circularly"
 & \\
 & &  & propagation.& propagation. & \\ \\
$ < {\mathop {S_ \oplus ^\phi  }\limits^{(1)} }
> _{\omega _e  = \omega _g } $ & $sin(2\phi)$ & 0 & $
| {< {\mathop {S_ \oplus ^\phi  }\limits^{(1)} }
> _{\omega _e  = \omega _g } } | = $ & $ |
{< {\mathop {S_ \oplus ^\phi  }\limits^{(1)} }
> _{\omega _e  = \omega _g } } | = $ & 0
\\
  & & & $  | {< {\mathop {S_ \oplus ^\phi  }\limits^{(1)} } > _{\omega _e  = \omega _g } } |_{\max
  },
$ & $  | {< {\mathop {S_ \oplus ^\phi }\limits^{(1)} } > _{\omega
_e  = \omega _g } }
|_{\max } $ ,\\
  & & & ``Left circularly"  & ``Right circularly"
 & \\
  & & & propagation.& propagation. & \\ \\
  $
{< {\mathop {S_ \otimes ^\phi  }\limits^{(1)} }
> _{\omega _e  = \omega _g } } $ & $sin^2 \phi $ & $
| {< {\mathop {S_ \otimes ^\phi  }\limits^{(1)} }
> _{\omega _e  = \omega _g } } | = $ & There are
& There are &  $ | {< {\mathop {S_ \otimes ^\phi }\limits^{(1)} }
> _{\omega _e  = \omega _g } } | =
$ \\
  & & $
| {< {\mathop {S_ \otimes ^\phi  }\limits^{(1)} }
> _{\omega _e  = \omega _g } } |_{\max } $
 & nonvanishing values, & nonvanishing values, &$
| {< {\mathop {S_ \otimes ^\phi  }\limits^{(1)} }
> _{\omega _e  = \omega _g } } |_{\max } $, \\
  & & at $\phi = \pi /2, 3 \pi /2$  & ``Left circularly"  &``Left circularly"
 &``Left circularly" \\
   & & & propagation.& propagation. & propagation. \\ \\
 \hline
\end{tabular}
\end{center}
\end{table}

Table I, Eqs.(60)-(63) and Figures 2, 3 show that the plane $z=0$
and the planes $\phi = \pi /2, \;\; 3\pi/2$ are the three most
interesting regions. For the former, $
 < \mathop {S^\phi  }\limits^{(0)}  >  =  < \mathop {S^r }\limits^{(0)}  >
  =  < \mathop {S_ \oplus ^\phi  }\limits^{(1)}  > _{\omega _e  = \omega _g }  \equiv 0
 $, but nonvanishing $
 < \mathop {S_ \otimes ^\phi  }\limits^{(1)}  > _{\omega _e  = \omega _g }
 $
exists; for the latter, $
 < \mathop {S^\phi  }\limits^{(0)}  >  =  < \mathop {S_ \oplus ^\phi  }\limits^{(0)}  > _{\omega _e  = \omega _g }  \equiv 0
 $, but there is non-zero $
 < \mathop {S_ \otimes ^\phi  }\limits^{(1)}  > _{\omega _e  = \omega _g }
$. This means that any nonvanishing tangential EM power flux in
such regions will express the pure electromagnetic-gravitational
perturbation.

\subsection{Numerical estimations}
If we describe the perturbation
in the quantum language (photon flux), the corresponding
perturbative photon flux  $n_\phi$ caused by $
 < \mathop {S_ \otimes ^\phi  }\limits^{(1)}  > _{\omega _e  = \omega _g }
 $ at the plane $\phi=\pi/2$ ({\it {we note
that $
 < \mathop {S_ \otimes ^\phi  }\limits^{(1)}  > _{\omega _e  = \omega _g }
 $ is the unique nonvanishing power flux density passing
through the plane}}) can be given by\\
\begin{equation}
n_\phi = \frac{{ < \mathop {u_ \otimes ^\phi  }\limits^{(1)}  >
_{\omega _e  = \omega _g ,\;\phi  = {\pi  \mathord{\left/
 {\vphantom {\pi  2}} \right.
 \kern-\nulldelimiterspace} 2}} }}{{\hbar \omega _e }}
= \frac{1}{{\hbar \omega _e }}\int_0^{W_0 } {\int_{ - l/2}^{l_0} {
< \mathop {S_ \otimes ^\phi  }\limits^{(1)}  > _{\omega _e  =
\omega _g ,\;\phi  = \pi /2} } } dzdr,
\end{equation}\\
where $
 < \mathop {u_ \otimes ^\phi  }\limits^{(1)}  > _{\omega _e  = \omega _g ,\;\phi  = {\pi  \mathord{\left/
 {\vphantom {\pi  2}} \right.
 \kern-\nulldelimiterspace} 2}}  = \int_0^{W_0 } {\int_{ - {l \mathord{\left/
 {\vphantom {l 2}} \right.
 \kern-\nulldelimiterspace} 2}}^{l_0 } { < \mathop {S_ \otimes ^\phi  }\limits^{(1)}  > _{\omega _e  = \omega _g ,\;\phi  = {\pi  \mathord{\left/
 {\vphantom {\pi  2}} \right.
 \kern-\nulldelimiterspace} 2}} } } dzdr$ is the total
 perturbative power flux passing through the plane $\phi=\pi/2$,
 $\hbar$ is the Planck constant.

In order to give reasonable estimation,we choose the achievable
values of the EM parameters in the present experiments:
(1)$\psi_0=3\times10^5V.m^{-1}$ (i. e., $\psi_0/c=10^{-3}T$), the
amplitude of the Gaussian beam. If the spot radius $W_0$ of the
Gaussian beam is limited in $0.05m$, the corresponding power can
be estimated as $ P = \int_0^{W_0 } { < \mathop {S^z
}\limits^{(0)}  > _{z = 0} 2\pi rdr \approx 10^5 W} $ (see Eq.
(9)), this power is well within the current technology condition
[21, 35]. For the Gaussian beam with $\nu_e=3\times10^{10}Hz$,
this is equivalent to a photon flux $n^{(0)}$ of
$3\times10^{28}s^{-1}$ roughly. (2)$ \hat B_y^{(0)} = 30T $, the
strength of the background static magnetic field, this is
achievable strength of a stationary magnetic field under the
present experimental condition [36]. (3) $A_ \otimes= 10^{-30}$,
$\omega_g/2\pi=\nu_g=3\times 10^{10}Hz$, these are the typical
orders expected by the quintessential inflationary models [1].
Substituting Eqs.(61),(63) and the above parameters into Eq.(65),
and setting $W_0=0.05m,l=0.1m,l_0=0.3m$,we obtain
$n_\phi\approx1.57\times10^3s^{-1}$ (see Table II). Up to now,
this is largest perturbative photon flux in a series of results
[7, 8, 9]. Recently, we analyzed and estimated them under the
typical laboratory parameter conditions. If the integration region
of the radial coordinate $r$ in Eq.(65) is moved to $W_0 \leq r
\leq r_0$ (here $W_0=0.05m$, $r_0=0.1m$), in the same way, the
corresponding perturbative photon flux $ n'_\phi $ can be
estimated as $ n'_\phi = \frac{1}{{\hbar \omega _e }}\int_{W_0
}^{r_0} {\int_{ - l/2}^{l_0 } { < \mathop {S^\phi _ \otimes
}\limits^{(1)}
> _{\omega _e  = \omega _g ,\phi  = \pi /2} dzdr = 0.96 \times
10^3 } } s^{ - 1} \approx 10^3 s^{ - 1} $. Although then $n'_\phi
< n_\phi $, it retains basically order of $10^{3}s^{-1}$, and
because the ``receiving" plane of the tangential perturbative
photon flux is already moved to the region outside the spot radius
$W_0$ of the Gaussian beam, it has more realistic meaning to
distinguish and display the perturbative photon flux.

\begin{figure}[htbp]
\centerline{\includegraphics[width=5.07in,height=5.17in]{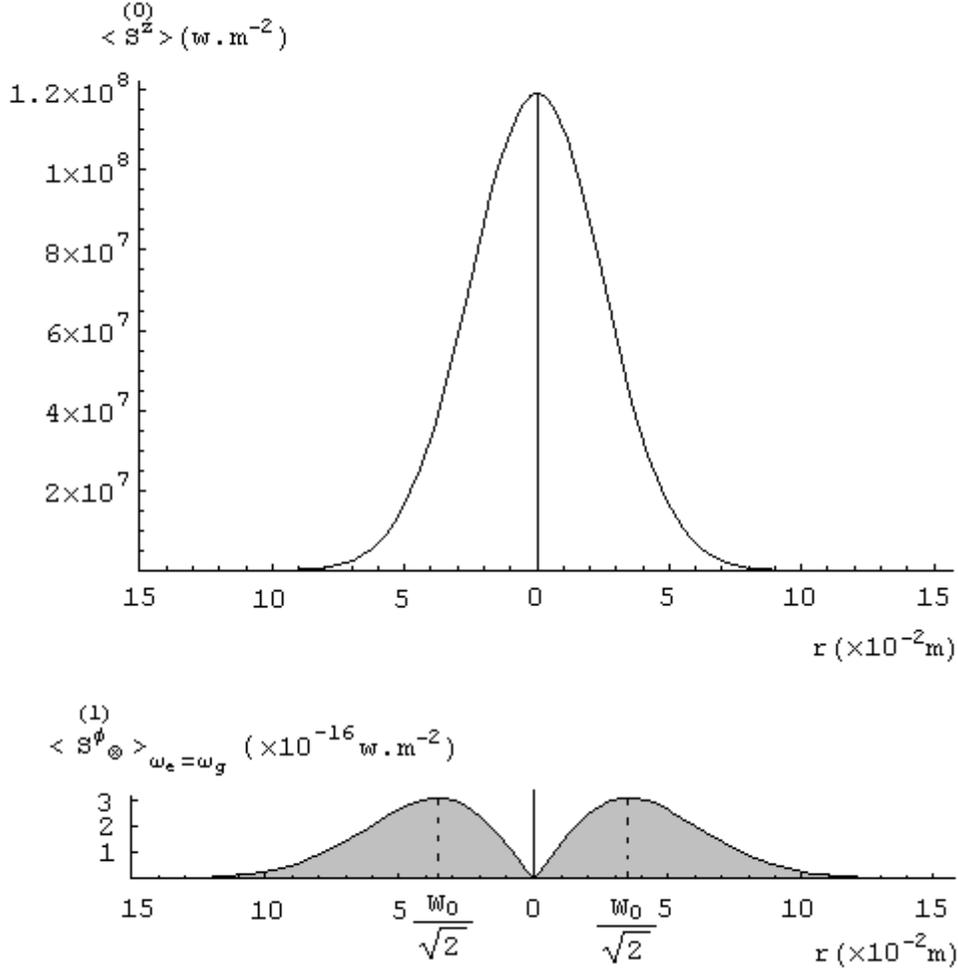}}
\caption{\footnotesize Rating curve between $
 < \mathop {S_ \otimes ^\phi  }\limits^{(1)}  > _{\omega _e  = \omega _g }
$ and $< \mathop {S^z }\limits^{(0)}  > $ at the plane $z=0$, hers
$r$ is the radial coordinate, and $\phi = \pi /2$ or $3 \pi /2 $
in Eq.(64). It shows that $ | < \mathop {S_ \otimes ^\phi
}\limits^{(1)}  > _{\omega _e  = \omega _g } |
 = | < \mathop {S_ \otimes ^\phi  }\limits^{(1)}  > _{\omega _e  = \omega _g } |_{\max }
$  at $r=W_0/ \sqrt 2 $. The background axial power flux density $
< \mathop {S^z }\limits^{(0)}  > $ is typical Gaussian
distribution, and $ | < \mathop {S^z }\limits^{(0)}  > | >  > | <
\mathop {S_ \otimes ^\phi  }\limits^{(1)}  > _{\omega _e  = \omega
_g } |_{\max } $. However, because the propagating direction of $<
\mathop {S^z }\limits^{(0)}  > $ is perpendicular to that of $<
\mathop {S_ \otimes ^\phi  }\limits^{(1)}  > _{\omega _e  = \omega
_g }$, $< \mathop {S^z }\limits^{(0)}  >$ has no any contribution
in the pure tangential direction. The shaded part expresses $<
\mathop {S_ \otimes ^\phi  }\limits^{(1)}  > _{\omega _e  = \omega
_g }$. Here $\nu_e = \nu _ g = \omega_g /2\pi = 3 \times 10 ^{10}
Hz$, $A_ \otimes = 10 ^{-30}$, $\hat B_y^{(0)} =30T$, $\psi_0 =
3\times10^5 V.m^{-1}$, $l=0.1m$, $W_0=0.05m $. }
\label{Vr_orbit1_part}
\end{figure}

Figure 4 gives the rating curve between $ < {\mathop {S_ \otimes
^\phi  }\limits^{(1)} } > _{\omega _e  = \omega _g } $ and $ <
{\mathop {S^z }\limits^{(0)} } > $ at the plane $z=0$, here $r$ is
the radial coordinate.

Figure 5 gives the rating curve between $n_\phi$ and the axis
coordinate $z$, and relative parameters are chosen as
$\nu_e=\nu_g=\omega_g/2\pi=3\times10^{10}Hz$,
$A_\otimes=10^{-30}$, $ \hat B_y^{(0)}  = 30T$,
$\psi_0=3\times10^5V.m^{-1}$, $l=0.1m$, $l_0=0.3m$ and
$W_0=0.05m$. Figure 5 shows that $n_\phi$ has a good space
accumulation effect as $z$ increases. In fact, in addition to the
third- and fourth-terms in Eq.(61), the rest in the expression of
$ < \mathop {S_ \otimes ^\phi  }\limits^{(1)}  > _{\omega _e  =
\omega _g }
 $, Eq.(61) and Eq.(63), is all slow
enough variational function at the $z$ direction. This means that
the value of $
 < \mathop {S_ \otimes ^\phi  }\limits^{(1)}  > _{\omega _e  = \omega _g } $ is
slow variational and keeps its sign invariant in the whole region
of the coherent resonance (here it is about the region of
$80\lambda_g$, namely $0.8m$), that they make $
 < \mathop {S_ \otimes ^\phi  }\limits^{(1)}  > _{\omega _e  = \omega _g } $ being all ``left
circularly" propagated from $-l/2$ to $l_o$.

\begin{figure}[htb]
\centering\noindent
\includegraphics[width=6.33in,height=3.83in]{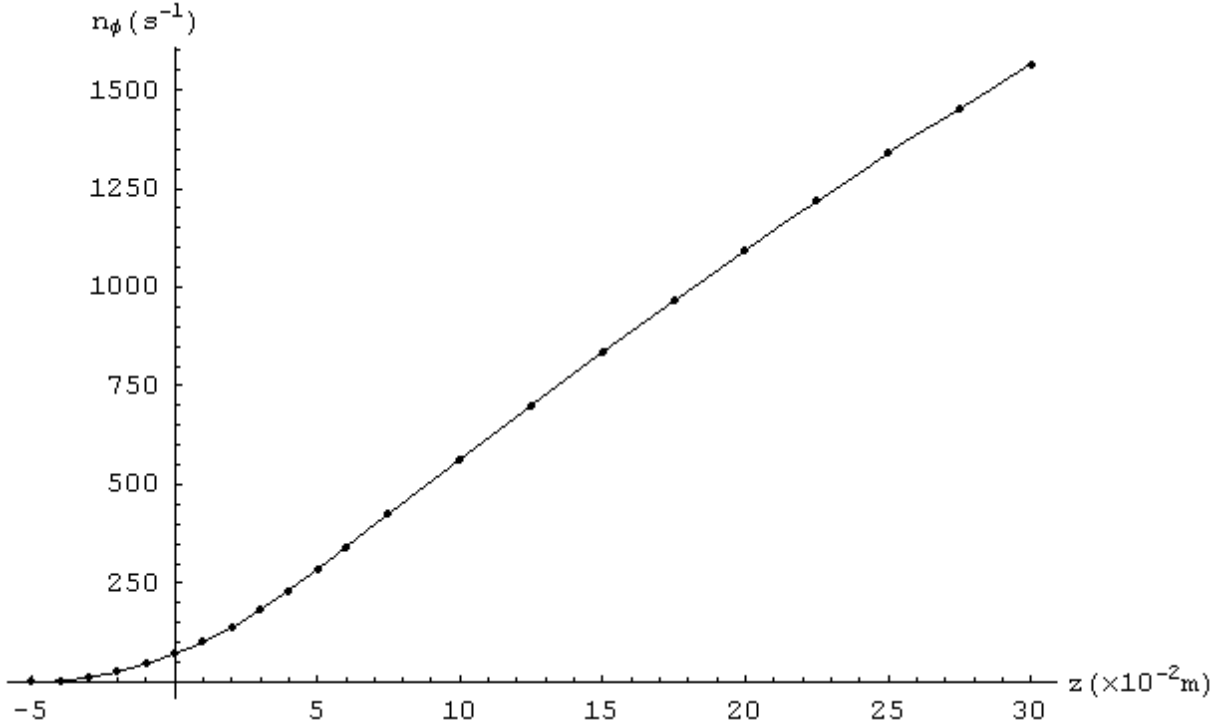}
\caption{\footnotesize Rating curve between $n_\phi$ and the axial
coordinate $z$, here $\nu_e=\nu_g=\omega_g/2\pi=3\times10^{10}Hz$,
$A_\otimes=10^{-30}$, $ \hat B_y^{(0)}  = 30T$,
$\psi_0=3\times10^5V.m^{-1}$, and $W_0=0.05m$. It shows that the
pure tangential perturbative photon flux passing through the plane
$\phi=\pi/2$ with $10^{-2}m^2$, would be expected to be $1.57
\times 10^3s^{-1}$. }
\label{Vr_orbit1_part}
\end{figure}

Table II gives the tangential perturbative photon fluxes and
corresponding relevant parameters in the three cases. In the first
case $W_{01}=0.05m$, in the second case $W_{02}=0.02m$, in the
third case $W_{03}=0.1m$, but the background Gaussian beam keep
the same power in such three cases, i. e., $P \approx 10^5W$. Due
to $W_{03}>W_{01}>W_{02}$, so that
$\psi_{03}<\psi_{01}<\psi_{02}$. Table II shows that the
tangential perturbation in the Gaussian beam with larger $W_0$ (i.
e., smaller $ \theta $) has a better physical effect than that in
the Gaussian beam with smaller $W_0$ (i. e., larger $\theta$),
here $\theta$ is the spreading angle of the beam,
$\theta=tan^{-1}( {\frac{{\lambda _e }}{{\pi W_0 }}} ) \approx
{\frac{{\lambda _e }}{{\pi W_0 }}}$\;.

\begin{table}[!htbp] 
\caption{The tangential perturbative photon fluxes  and
corresponding relevant parameters }
\begin{center}
\begin{tabular}{lcccccc}\hline
{\;\;\;\;\;  } & {$A$} & {$\nu _ g (Hz)$} & {$W_0 (m)$} &
{$\theta$} & {$ < {\mathop {u_ \otimes ^\phi  }\limits^{(1)} }
> _{\omega _e  = \omega _g } $} (\it {W}) & {$n_\phi
(s^{-1})$}\\ \hline  \\
(1)  & $10^{-30}$ & $3\times 10^{10}$  & $\;\;\;\;\;0.05\;\;\;\;\;\;$
& $6.36\times10^{-2}$  & $\;\;\;\;3.11\times10^{-20}\;\;\;\;$  & $\;\;\;1.57\times10^3\;\;\;$\\
\\ (2)  & $10^{-30}$ & $3\times 10^{10}$  & $0.02$ &$15.78\times 10^{-2}$&$ 1.31\times10^{-20}$ &$6.61\times10^2$\\
\\(3) &$\;\;\;\;10^{-30}\;\;\;\;\;$&$3\times10^{10}$& $0.10$ & $3.18\times
10^{-2}$& $3.14\times10^{-20} $  & $1.58 \times 10^3$
\\  \\ \hline
\end{tabular}
\end{center}
\end{table}

We emphasize that here $ n_\phi   \propto \psi _0  \propto \sqrt
P$ (see Eqs.(61), (63), and (65), $P$ is the power of the Gaussian
beam), and at the same time, $n_\phi \propto \hat B_y^{(0)} $.
Therefore, if $P$ is reduced to $10^3 W$, then $n_\phi\approx
1.57\times 10^2 s^{-1}$; and even if $P$ is reduced to $10 W$
(this is already a very relaxed requirement), we have still
$n_\phi \approx 1.57 \times 10 s^{-1}$. Thus, if possible,
increasing $\hat B_y^{(0)}$ (in this way the number of the
background real photons does not change) has better physical
effect than increasing $P$. According to the above discussion, we
give some values for the power $P$ of the background Gaussian beam
and corresponding parameters $n_\phi$ and $ n'_\phi $ (see Table
III).

\begin{table}[!htbp] 
\caption{The power of the background Gaussian beam and
corresponding $n_\phi$ and $n'_\phi$.}
\begin{center}
\begin{tabular}{ccc}\hline
$\;\;\;\;\;\;\;\;\;\;P(W)\;\;\;\;\;\;\;\;\;\;$ &
$\;\;\;\;\;\;\;\;\;\; n_\phi(s^{-1})\;\;\;\;\;\;\;\;\;\;$ &
$\;\;\;\;\;\;\;\;\;\; n'_\phi(s^{-1})\;\;\;\;\;\;\;\;\;\;$ \\ \hline  \\
 $10^5$ & $\sim 1.57\times 10^3$  & $\sim 10^3$\\
\\  $10^{3}$ & $\sim 1.57\times 10^2$ &$\sim 10^2$\\
\\ $10$ & $\sim 1.57 \times 10$ & $\sim 10$
\\  \\ \hline
\end{tabular}
\end{center}
\end{table}

In particular, since $n'_\phi$ indicates the tangential
perturbative photon flux passing through the ``receiving" plane
$\phi=\pi/2$ ($\sim 10^{-2}m^2$) outside the spot radius $W_0$ of
the Gaussian beam, the results provided a more realistic test
scheme.

\section{GEOMETRICAL PHASE SHIFT PRODUCED BY THE HIGH-FREQUENCY RELIC GRAVITATIONAL WAVES}
$\;\;\;\;$Mitskievich et al. [37] investigated Berry's phase shift
of a monochromatic EM wave beam in a plane monochromatic GW field,
and it is shown that (a) For parallel the propagating GW and EM
wave, the phase shift is absent. (b) When the both waves are
mutually orthogonal, the nonvanishing phase shift will be
produced, and the phase shift is proportional to the distance
propagated by the EM wave (see, Eq.(6) in Ref. [37]). i. e.,\\
\begin{equation}
\Delta \alpha  = A\left( {\Lambda  - \sin \Lambda } \right)\sin
\left( {\Lambda  + \delta '} \right),
\end{equation}\\
where\\
\begin{equation}
\Lambda  = 2\pi L/\lambda _g = \frac{L}{c} \omega _g,
\end{equation}\\
A is the amplitude of the GW (e. g., $A_\oplus $ or $A_ \otimes$),
$L$ is the distance between the observer and reflecting system of
the EM wave, and $\lambda_e = \lambda _g,\omega _e = \omega _g$,
$L$ is also the interacting dimension of the GW with the EM wave.
If $\Lambda >> 1$, from Eq.(66), we have\\
\begin{equation}
\Delta \alpha  \approx A \Lambda \sin \left( {\Lambda  + \delta '}
\right),
\end{equation}\\
namely, one obtained a linear increase of the phase shift.
Eqs.(67) and (68) show that large distance $L$ and high-frequency
$\omega _g $ will produce better physical effect than that of the
small $L$ and low frequency $\omega _g$.

As we have pointed out that unlike an ideal plane monochromatic EM
wave, the Gaussian beam is a realized EM wave beam satisfying
physical boundary conditions, and as we shall show that for a
Gaussian beam with a small the spreading angle, the wave beam in
the region near the symmetrical axis can be approximately seen as
a quasi-plane wave. In this case we are possible to estimate the
geometrical phase shift in the Gaussian beam.

In order to simplify our analysis, we consider only the real part
of Eq.(1). From Eq.(1) we have\\
\begin{equation}
Re(\psi) = f_1(r,z)cos(k_e z-\omega _e t) + f_2 (r,z)sin (k_e
z-\omega _e t),
\end{equation}\\
where\\
\begin{equation}
f_1 (r,z) = \frac{{\psi _0 }}{{\left[ {1 + \left( {{z
\mathord{\left/
 {\vphantom {z f}} \right.
 \kern-\nulldelimiterspace} f}} \right)^2 } \right]^{1/2} }}
 \exp(-\frac {r^2}{W^2}) \cos \left( {\tan ^{ - 1} \frac{z}{f} - \frac{{k_g r^2 }}{{2R}} - \delta }
 \right),
\end{equation}\\
\begin{equation}
f_2 (r,z) = \frac{{\psi _0 }}{{\left[ {1 + \left( {{z
\mathord{\left/
 {\vphantom {z f}} \right.
 \kern-\nulldelimiterspace} f}} \right)^2 } \right]^{1/2} }}
 \exp(-\frac {r^2}{W^2}) \sin \left( {\tan ^{ - 1} \frac{z}{f} - \frac{{k_g r^2 }}{{2R}} - \delta }
 \right).
\end{equation}\\
Eqs.(69)-(71) show that for the Gaussian beam with the small
spreading angle, deviation of the propagation direction from the
$z$-axis in the region near axis would be very small, and for the
high-frequency band, the functions $f_1$ and $f_2$, Eqs.(70) and
(71), will be slow variational functions at the $z$ direction. In
this case the change of the Gaussian beam as space-time mainly
depends on the propagation factors $cos(k_e z - \omega _ e t)$ and
$sin(k_e z - \omega _e t)$. In this sense it is just
characteristic of the plane wave. Therefor, for the GW propagating
along the $y$-axis, because its propagating direction is
perpendicular  to the $z$-axis, it would be to generate the phase
shift satisfying approximately Eq.(66) or Eq.(68). Notice that
since in this case the propagating direction of the GW is parallel
with the static magnetic field $ \hat B_y^{(0)} $ , whether from
the classical or the quantum theories of the weak fields, the GW
(gravitons) does not produce perturbative EM fields (photon fluxes
) to the static magnetic fields [28, 29]. Thus then the all
first-order perturbations expressed as Eqs.(54)-(56) vanish.

It is interesting to compare the geometrical phase shift produced
by the high-frequency relic GW of $\nu _ g = 3 \times 10^{10}Hz$
and $A=10^{-30}$ in the Gaussian beam with the phase shift
generates by the expected astronomical GW of $\nu _ g = 3\times
10^3 Hz$ and $A=10^{-22}$ in the plane monochromatic EM wave (see
Table IV). With the help of Eqs.(66)-(68), we list the some
typical parameters in Table IV.

\begin{table}[!htbp] 
\caption{The geometrical phase shift produced by the GW's in the
EM wave beams with the same frequencies}
\begin{center}
\begin{tabular}{lcccccc}\hline
{  } &{$A$} & {$\nu _ g = \nu _ e (Hz)$} & {The wave form} & { The interacting } & {$\Lambda$} & {The geometrical }\\
{  } & { } & { } & {} & {dimensions $L$ (m)} & { } & {phase shift
$\Delta \alpha$} \\
 \hline \\
(a) & $\;\;\;\;\;10^{-22}\;\;\;\;\;$ & $3\times 10^3$ &
\;\;Monochromatic\;\; & $3.8 \times 10^8 $ & $\;\;\;2.4 \times
10^4\;\;\;$ & $\;\;\;2.4 \times 10^{-18}\;\;\;$
\\
  &  &  &  plane EM wave  & \;\;(Cislunar distance)\;\; \\ \\
(b) & $10^{-30}$ & $3\times 10^{10}$ &  Gaussian beam  & $1$ &
$2.5 \times 10^2$ & $2.5 \times 10^{-28}$\\ \\
(c) &$10^{-30}$ &  $3\times 10^{10}$  & Gaussian beam & 38 & $2.4
\times 10^4$  & $2.4 \times 10^{-26}$\\ \\
(d) & $10^{-30}$ &  $3\times 10^{10}$  & Gaussian beam & $5\times 10^9$ & $3.1 \times 10^{12}$ & $3.1\times 10^{-18}$\\
& & & & (LISA dimension) \\ \hline
\end{tabular}
\end{center}
\end{table}

We can see from Table IV that scheme (a) has typical parameters of
the expected astronomical GW's [see analysis in Ref.[37] and
Eqs.(60)-(68)], and $\Lambda = 2.5 \times 10^4$, $\Delta \alpha =
2.4\times 10 ^{-18}$. This means that in this case a GW of
$A=10^{-22}$ and $\nu _ g = 10^3Hz$ can be treated as an effective
magnitude of some $10^{-18}$, but it needs the interacting
dimension of cislunar distance (i. e., the reflecting system of
the EM wave beam is placed on the surface of moon). For the scheme
(d), the phase shift may achieve the same order of magnitude with
the scheme (a), but it needs a  interacting dimension of the LISA
($\sim10 ^9m$), and it is necessary to constructing a very strong
Gaussian beam with the small spreading angle in microwave
frequency band for the LISA. This seems to be beyond the ability
of presently conceived technology. Nevertheless, if the amplitude
of the high-frequency relic GW ( $\nu _ g \sim 10^{10}Hz$) and the
amplitude of the expected astronomical GW ($\nu _g \sim 10^3Hz$)
have the same order of magnitude, then the geometrical phase shift
$\Delta \alpha$ produced by the former will be seven orders larger
than that generated by the latter.

\section{CONCLUDING REMARKS}
1.\;\;For the relic GW's expected by the quintessential
inflationary models, since a large amount of the energy of the
GW's may be stored around the GHz band, using smaller EM systems
(e. g., the microwave cavities or the Gaussian beam discussed in
this paper) for detection purposes seems more plausible. In
particular the Gaussian beams can be considered as new possible
candidates. For the high-frequency relic GW with typical order of
$\nu _ g = 10^{10}Hz$, $h=10^{-30}$ in the models, under the
condition of resonant response, the corresponding first-order
perturbative photon flux passing through the region $10^{-2}m$
would be expected to be $10^{3}s^{-1}$. This is the largest
perturbative photon flux we have recently analyzed and estimated
using the typical laboratory parameters.

2.\;\;In our EM system, the perturbative effects produced by the +
and $\times$ polarization states of the high-frequency relic GW
have a different physical behavior. Especially, since the
first-order tangential perturbative photon flux produced by the
$\times$ polarization state of the relic GW is perpendicular to
the background photon fluxes, it will be unique nonvanishing
photon flux passing through the some special planes. Therefore any
photon measured from such photon flux at the above special planes
may be a signal of the EM perturbation produced by the GW, this
property may be promising to improve further the EM response to
the GW.

3.\;\;As for the geometrical phase shift produced by the
high-frequency relic GW, because of the excessive small amplitude
orders of the relic GW, the phase shift is still below the
requirement of the experimental observation. Thus the outlook for
such schemes may be not promising unless there are stronger
high-frequency relic GW's. But for the dimensions of the LISA (of
course, in this case it needs a very strong microwave beam), it is
possible to get an observable effect.

The relic GW's are quite possible the few windows from which we
can look back the early history of our universe, while the
high-frequency relic GW's in the GHz band expected by the
quintessential inflationary models, are possible to provide a new
criterion to distinguish the quintessential inflationary and the
ordinary inflationary models. As pointed out by Ostriker and
Steinhardt [5], whatever the origin of quintessence, its dynamism
could solve the thorny problem of fine-tuning our universe. If we
could display the signal of the high-frequency relic GW's in the
quintessential inflationary models through the EM response or
other means, it would not only previde incontrovertible evidence
of the GW and quintessence, but also give us the extraordinary
opportunity to look back the early universe. Therefore, there is
even a small chance that the signal of the high-frequency relic
GW's is detectable, then it is worth pursuing.

Finally, it should be pointed out that the relic GW's are the
stochastic backgrounds, their signals have often larger
uncertainty. However, it can be shown that for the relic GW's
spectra expressed as Eq. (12), only one component (e.g. the form
expressed as Eq. (15)) with special propagation direction,
frequency, polarization and phase, can produce maximal
perturbation to our EM system. Therefore, the high-frequency relic
GW's would be instantaneously detectable at least. Moreover, if we
can find a very subtle means in which the pure perturbative photon
flux can be pumped out from our EM system, so that the
perturbative and background photon flux can be completely
separated, then the detection possibility of the relic GW's would
be greatly increased. The more detail investigation will be
discussed elsewhere.

\section{Acknowledgments}
One of the authors (F. Y. Li) would like to thank J. D. Fan, Y. M.
Malozovsky, W. Johnson and E. Daw for very useful discussions and
suggestions. This work is supported by the National Nature Science
Foundation of China under Grants No. 10175096, No. 19835040 and by
the Foundation of Gravitational and Quantum Laboratory of Hupeh
Province under Grant No. GQ0101.

\appendix
\section{\;THE CAVITY ELECTROMAGNETIC RESPONSE TO THE GRAVITATIONAL WAVES}
For the cavity electromagnetic response to the GW's, the best
detecting state is the resonant response of the fundamental EM
normal modes of the cavity to the GW's, since in this case it is
possible to generate the maximal EM perturbation. Under the
resonant states (whether the background EM fields stored inside
the cavity is only a static field or both the static magnetic
field and the normal modes), the displaying condition at the level
of the quantum nondemolition measurement can be written as [10, 11]\\
\begin{eqnarray}
\frac{(hQ)^2B^2V}{\mu_0 \hbar \omega_e}\geq 1,
\end{eqnarray}\\
where $Q$ is the quality factor of the cavity, $V$ is its volume,
$B$ is the background static magnetic field. In order to satisfy
the fundamental resonant condition, the estimated cavity
dimensions should be comparable to the wavelength of the GW.
However, the dimensions cannot be too big. This is because
constructing a superconducting cavity with typical dimension
$l\geq100m$ may be unrealistic under the present experimental
conditions. The low-frequency nature of the usual astronomical
GW's seem to greatly limit the perturbative effects in the
cavity's fundamental EM normal modes. For the high-frequency GW's
in GHz band, the corresponding resonant condition can be relaxed,
but the cavity size cannot be excessive small, even if the
condition $l \geq \lambda_g $ can be satisfied, since in this case
the cavity cannot store enough EM energy to generate observable
perturbation. If GW's detected by the cavity have excessive
high-frequency, e. g., $\nu_g=\nu_e>10^9 Hz$, we can see from
Eq.(A1) that the requirements of other parameters will be a big
challenge. For instance, if one hopes to detect the high-frequency
relic GW with $h=10^{-30}$ and $\nu_g=10^9 Hz$ in the
quintessential inflationary models, we need $Q=10^{12}$, $B=30T$
and $V=100m^3$ at least, then the corresponding signal
accumulation time will be $\tau \approx Q/\omega_e\approx 10^3 s
$. If $h=10^{-30}$, $\nu_g=10^8Hz$, we need $Q=10^{12}$, $B=30T$
and $V=10m^3$ (i. e., the typical dimension of the cavity will be
$l\sim2.2m$) at least, then $\tau \approx Q/\omega_e\approx 10^4 s
$. Increasing quality factor $Q$ and using squeezed quantum states
may be a promising direction [11]. For the former, the
requirements of other parameters can be further relaxed; for the
latter, the signal accumulation time could be decreased.
Therefore, for the fundamental resonant response, the suitable
size of the cavity may be the magnitude of meter, the
corresponding resonant frequency band should be
$10^8Hz<\nu_g<10^9Hz$ roughly (the region (4)-2 in Figure I). In
order to detect the GW's of $10^8Hz<\nu_g<10^9Hz$, the thermal
noise must be $KT<h\nu_g$ which corresponds to $T \leq 10^{-3}K$
(see Appendix D).

Moreover, the resonant schemes of difference-frequency suggested
in Refs.[38, 39] can be used as EM detectors of the high-frequency
GW's. The detector consists of two identical high-frequency
cavities (e. g., two coupled spherical cavities discussed in
resent paper [39]). When the GW frequency $\nu_g$ equals the
frequency difference $|\nu_1-\nu_2|$ of the two cavity modes (i.
e., $\nu_g=|\nu_1-\nu_2|$, and $\nu_1$, $\nu_2\gg \nu_g$), then
detector can get maximal EM energy transfer. Following [39] we can
learn that sensitivity of the EM detector would be expected to be
$\delta h \sim 10^{-20}-10^{-22}$ for the GW's of $10^3-10^4Hz$
frequency band (the region (4)-1 in Figure I). If this EM detector
is advanced, it might to detect the GW's in GHz band. Ref. [40]
reported an EM detecting scheme of the high-frequency GW's by the
interaction between a GW and the polarization vector of an EM wave
in repeated circuits of a closes loop. In the scheme because of
 linearly cumulative effect of the rotation of the polarization
 vector of the EM wave, expected sensitivity can reach up to
 $\delta h \sim 10^{-18}-10^{-19}$ for the GW's of $10^8-10^9Hz$ (the region (4)-2 in Figure I), and the above two
 scheme [39, 40] are all sensitive to the polarization of the
 incoming GW signal. Although the sensitivity of the above EM
 detectors is still below the requirements of the observable effect
 to the high-frequency relic GW's, the advanced EM detector
 schemes would be promising.

 \section{\;THE DIMENSIONLESS AMPLITUDE $\;h\;$ AND THE POWER SPECTRUM
$\;S_h\;$ OF THE HIGH-FREQUENCY RELIC GRAVITATIONAL WAVES} The
relation between the logarithmic energy spectrum $\Omega_{GW}$ and
the power spectrum $S_h$ of the relic GW given by Ref.[1] (see
Eq.(A18) in Ref.[1]) is
\begin{eqnarray}
\Omega_{GW}(\nu,\eta_0)=\frac{4\pi^2}{3H_0^2}\nu_g^3S_h(\nu,\eta_0),
\end{eqnarray}
where $H_0$ is the present value of the Hubble constant, i.e.,
$H_0=3.24\times 10^{-18} s^{-1}$. From Eq.(B1), one finds [1]
\begin{eqnarray}
S_h(\nu,\eta_0)\approx 8\times 10^{-37}
\Omega_{GW}({\nu,\eta_0})\frac{(Hz)^2}{\nu_g^3}.
\end{eqnarray}

 In the peak region of the logarithmic energy spectrum of the relic GW in the
quintessential inflationary models, $\Omega_{GW}\approx 5 \times
10^{-6}$[1]. Thus, for the relic GW of $\nu=10^9 Hz$, we have $S_h
\approx 4 \times 10^{-69}s$; for the relic GW of $\nu=10^{10} Hz$,
one finds $S_h \approx 4 \times 10^{-72}s$.

  For the continuous GW's, the dimensionless amplitude $h$ can be
estimated roughly as
\begin{eqnarray}
h\approx (S_h \Delta\nu)^\frac{1}{2},
\end{eqnarray}
 where $\Delta\nu_g$ is
corresponding bandwidth. According to estimation in Ref.[1] (see
Figure.2 in Ref.[1] or see Figure.1 in this paper), the bandwidth
in the high-frequency peak region is about $\Delta\nu\approx
10^{11} - 10^9 Hz\approx10^{11} Hz$. Thus, from Eq.(B3), we have
\begin{eqnarray}
h\approx 10^{-29}\sim 10^{-30}\;\;\; (\rm{for\; the\; relic\; GW\;
of\;} \nu=10^9 Hz),\nonumber
\end{eqnarray}
\begin{eqnarray}
h\approx 10^{-30}\sim 10^{-31}\;\;\; (\rm{for\; the\; relic\; GW\;
of\;} \nu=10^{10} Hz).
\end{eqnarray}

The above results and orders of magnetude estimated in Ref.[1] are
basically consistent. In this paper we have chosen $h\sim
10^{-30}$. Of course, these estimations are only approximate
average effects. In fact, because of uncertainty of some relative
cosmological parameters [1, 2] in certain regions, it is possible
to cause small deviation to the above estimations.

\section{MINI-ASTRODYNAMICAL SPACE TEST OF RELATIVITY USING
OPTICAL DEVICES (Mini-ASTROD)} Mini-ASTROD is a new cooperation
project (China, Germany, etc.)[26]. The basic scheme of the
Mini-ASTROD is to use two-way laser interferometric ranging and
laser pulse ranging between the Mini-ASTROD spacecraft in
solar-system and deep space laser stations on Earth to improve the
precision of solar-system dynamics, solar-system constants and
ephemeris, to measure the relativistic gravity effects and to test
the fundamental laws of space-time more precisely, to improve the
measurement of the time rate of change of the gravitational
constant, and to detect low-frequency GW's ($\sim
10^{-6}Hz-10^{-3}Hz$).

  The follow-up scheme of the Mini-ASTROD is ASTROD[41], i.e., the
Mini-ASTROD is a down-scaled version of the ASTROD. Both LISA and
Mini-ASTROD are all space detection projects, but there are some
differences for their study objectives, detecting frequency band
of the GW's for the Mini-ASTROD will be moved to $10^{-6}Hz$ (see
Fig.1). With optical method, Mini-ASTROD should achieve the same
sensitivity as LISA [26]. Thus, Mini-ASTROD [26], ASTROD [41] and
LISA have a certain complementarity.

\section{NOISE PROBLEMS}
The noise problems of the EM detecting systems have been
extensively discussed and reviewd [1, 10, 35, 38, 39], here we
will give only a brief review to relevant problems of the EM
detection of the high-frequency GW's (especially the EM response
of the Gaussian beam).

 The thermal noise is one of the fundamental
sourse of limitation of the detecting sensitivity[38, 39]. Unlike
the usual mechanical detectors, the resonant frequencies of the EM
systems to the high-frequency GW's in $GHz$ band are often much
higher than ones of usual environment noises (e.g., mechanical,
seismic and others). Thus the EM detecting systems are easier to
reduce or shield external EM noise (e.g., using Faraday cages)
than closing out mechanical vibration from a detection system. For
the EM response of the microwave cavities to the high-frequency
GW's, the noise problem can be more conveniently treated
considering the relevant quantum character [35]. For a
superconducting cavity at a temperature $T=T_0$, if the background
EM field is only a static magnetic or static electric field, the
displaying condition can be given by Eq.(A1), while then the
cavity vacuum contains thermal photons with an energy spectrum
 given by the Plank formula:
\begin{eqnarray}
u_\nu  \left( \nu  \right) = \frac{{8\pi \nu ^2 }}{{c^3
}}\frac{{h\nu }}{{\exp \left( {\frac{{h\nu }}{{KT_0 }}} \right) -
1}},
\end{eqnarray}\\
where $u_\nu$ and $\nu$ are energy density and the photon
frequency respectively, while $K$ is the Boltzmann constant. If
the cavity is cooled down $T_0=1mK$, according to the Wien law,
the energy density has a maximum at $\nu_m = 5.87\times10 ^7 Hz$
(i.e., $\lambda_m \approx 3m$, corresponding photon density is
about $10^{-8}cm^{-3}$). For the perturbative photons produced by
the high-frequency GW of $\nu_g=3\times10^9Hz$ under the resonant
condition, we have $\nu_e=\nu_g$ (i.e., $\lambda=0.1m$), which is
higher than $\nu_m$ (i.e., $\nu_e \approx 30\nu_m$). Therefor, the
crucial parameter for the thermal noise is the selected frequency
and not the total background photon number. Namely, in this case
the thermal noise can be effectively suppressed as long as the
detector can select the right frequency.

For the EM response of the Gaussian beam in high-frequency region
of $\nu_e = \nu_g = 3 \times 10^{10}Hz$, because the frequencies
of usual environment noise are much lower than $\nu _ e$, the
requirements of suppressing such noise can be further relaxed.
While for the possible external EM noise sources, using a Faraday
cage would be very useful. Once the EM system (the Gaussian beam
and the static magnetic field) is isolated from the outside world
by the Faraday cage, possible noise sources would be the remained
thermal photons and self-background action. However, since the
``random motion" of the remained photons and the specific
distribution of the photon fluxes in the EM system (as we have
discussed earlier), the influence of such noise to the highly
``directional" propagated perturbative photon fluxes would be
effectively suppressed in the local regions. Therefore, the key
parameters for the noise problems are the selected perturbative
photon fluxes (e.g., $n_\phi$ and $n'_\phi$) passing through the
special planes (e.g., planes $\phi=\pi/2$ and $3\pi/2$) and not
the all background photons. Moreover, low-temperature vacuum
operation might effectively reduce the frequency of the remained
thermal photons and avoid dielectric dissipation. If the frequency
$\nu_m$ of the remained thermal photons is much lower than that of
the perturbative photon fluxes, i.e., $\nu_m\ll \nu_e$, then the
above two kind of photons would be more easily distinguished.

\end{document}